\RequirePackage{fix-cm}

\documentclass{article}

\usepackage{graphicx}
\usepackage{amssymb}
\usepackage{amsmath}
\usepackage{eucal}

\usepackage[driverfallback=dvipdfm,breaklinks=true,pdfcreator={LaTeX e Dvips}]{hyperref}
\usepackage{float}

\usepackage{mathptmx}
\usepackage[numbers]{natbib}
\usepackage{color}
\usepackage{ulem}
\usepackage{booktabs}
\usepackage{multirow}
\usepackage{textcomp}
\usepackage{bm}

\makeatletter
\let\MYcaption\@makecaption
\makeatother
\usepackage{subcaption}
      \makeatletter
\let\@makecaption\MYcaption
\makeatother

\usepackage[ruled,vlined]{algorithm2e}

\usepackage{placeins}

\title{Topology optimization of passively moving rigid bodies in unsteady flows}

\author{Yuta Tanabe$^\text{a,}\footnote{Corresponding author: {\tt 4524702@ed.tus.ac.jp} (Yuta Tanabe)}$,
        Kentaro Yaji$^\text{b}$ ,
        Kuniharu Ushijima$^\text{a}$ \\[12pt]
$^\text{a}$\textit{Department of Mechanical Engineering, Tokyo University of Science,}\\
              \textit{6-3-1, Niijuku,
              Katsushika-ku, Tokyo 125-8585, Japan}\\
$^\text{b}$\textit{Department of Mechanical Engineering, The University of Osaka,}\\
              \textit{2-1, Yamadaoka,
              Suita, Osaka 565-0871, Japan}}

\begin{document}

\maketitle

\begin{abstract}
    This study proposes the topology optimization method for moving rigid bodies subjected to forces from fluid flow, such as sails and turbines, with an unsteady time-dependent formulation.
    Unlike existing topology optimization frameworks in which rigid-body motion drives the flow, which is referred to as \textit{active}, the present study considers rigid-body motion induced by fluid forces, i.e., \textit{passive}.
    The equations of motion governing the rigid-body dynamics are solved in a coupled manner with the continuity equation and the momentum conservation equations.
    The rigid body is represented on a design grid that is separated from the analysis grid on which the state and adjoint fields are defined.
    After updating the rigid body motion, the body is mapped onto the analysis grid.
    The fluid equations are solved using the lattice kinetic scheme, an extended version of the lattice Boltzmann method, owing to its suitability for unsteady flows.
    Design sensitivities based on the adjoint variable method are presented and applied to two- and three-dimensional problems involving translational and rotational motions.
    The optimized shapes for each problem are discussed from a physical perspective and compared with a reference shape or their binarized counterparts, providing insights into the effectiveness of the proposed method as well as its limitations.
    \flushleft
    \textbf{Keywords}\ \ Topology optimization $\cdot$ Fluid flow $\cdot$ Unsteady problem $\cdot$ Moving rigid body $\cdot$ Separated design-analysis grids $\cdot$ Lattice Kinetic Scheme
\end{abstract}

\section{Introduction}
\label{sec1}

There are many kinds of devices that operate by forces generated from fluid flows, such as sails and turbines.
The efficiency of such devices depends on their shapes; therefore, optimizing their shapes is of significant interest.
In this study, we propose a topology optimization method for rigid bodies driven by fluid-flow-induced forces.

Topology optimization is one of the most powerful methods for obtaining shapes optimized with respect to a given objective function under some constraint functions.
This method was originally developed in the field of structural mechanics by \cite{bendsoe1988generating} and was later extended to fluid problems by \cite{borrvall2003topology}.
Although their work~\citep{borrvall2003topology} addressed only Stokes flow, extensive studies have subsequently been conducted for various fluid problems, including laminar steady Navier–Stokes flows~\citep{gersborg2005topology,Olesen2006}, turbulent flows~\citep{kontoleontos2013adjoint,DILGEN2018363}, unsteady flows~\citep{kreissl2011topology,DENG20116688}, and non-Newtonian flows~\citep{PINGEN20102340,10.1007/s00158-015-1346-5}.
Within this context, by considering a variety of objective and constraint functions, topology optimization has been applied to a wide range of engineering designs, such as airfoils~\citep{Kondoh2012,Li2022,Ghasemi2022}, fluid diodes~\citep{10.1115/1.4030297,Sasaki2024}, and microfluidic mixers~\citep{https://doi.org/10.1002/fld.1964,DONG2022131367}.
A comprehensive review of existing studies on fluid topology optimization can be found in \cite{fluids5010029}.

Although the applications mentioned above concern stationary components, topology optimization has also been applied to moving components, such as rotors~\citep{ROMERO2014268,OKUBO202116} and screw-type pumps~\citep{ALONSO2025115982}.
However, these studies mainly focus on devices whose motion is prescribed a priori; in other words, devices that drive fluid flow rather than devices driven by it.
Several works have examined the motion of particles~\citep{Andreasen2020,YOON2020113096,Yoon2021,CHEN2025106573} as moving entities driven by fluid flow.
In these studies, the objective functions are defined in terms of particle motion, such as travel distance or final position.
However, the optimization targets are stationary fluid channels, and the particles do not affect the fluid flow—i.e., a one-way coupling.
In contrast, this study focuses on optimizing the shape of a rigid body that is itself driven by fluid-flow forces.
The governing equations for fluid flow are solved in conjunction with the rigid-body equation of motion.
The forces appearing in the equation of motion are computed from the fluid flow, and the objective function is defined in terms of the rigid-body motion, such as displacement or rotation.

A closely related field is fluid–structure interaction (FSI), in which the governing equations for fluid flow are coupled with those for structural deformation.
The pioneering work on the topology optimization for FSI problems was conducted by \cite{https://doi.org/10.1002/nme.2777}.
Following this initial study, topology optimization for FSI has been extensively investigated using various approaches~\citep{Jenkins2015, Jenkins2016, PICELLI201744, Picelli2020, FEPPON2020109574, LI2022276} and problem settings~\citep{Lundgaard2018}.
While early studies considered only infinitesimal deformations and were restricted to steady-state fluid flow, more recent works have addressed large deformations~\citep{Silva2022} and unsteady fluid flows~\citep{YOON2023115729}.
However, the primary focus of that research field is structural deformation; when locomotion is treated, the objective is typically to minimize it~\citep{Lundgaard2018}, whereas the present study focuses on maximization of locomotion.
Consequently, typical benchmark problems in those studies involve cantilever beams or walls subjected to fluid forces.

The core technique enabling the treatment of moving rigid bodies subjected to fluid-flow forces is the grid separation approach~\citep{TANABE2025114620}.
In this approach, the design grid, where the shape is defined, is separated from the analysis grid, where state variables such as fluid velocity and pressure are computed.
At each time step, the design grid undergoes rigid-body motion and is then overlapped on the analysis grid.
Quantities such as the Brinkman coefficient and the velocity of points on the rigid body are transferred from the design grid to the analysis grid.
Because this approach directly represents rigid-body motion, it is well suited for problems involving rigid bodies driven by fluid flow.
Notably, the key difference between this work and the previous one~\citep{TANABE2025114620} lies in the treatment of the rigid-body motion.
In the previous work, the rigid-body motion is \textit{active}, that is, the motion is prescribed and drives the fluid flow.
In contrast, in this work, the motion is \textit{passive}, such that the rigid-body motion is fully coupled with the fluid flow, and the motion variables, including the position and velocity of the rigid body, are treated as additional state fields.
Rigid-body motion and fluid flow are solved simultaneously as an unsteady coupled problem.

Fluid flow is computed using the Lattice Kinetic Scheme (LKS)~\citep{Coveney2002lattice}, an extended version of the Lattice Boltzmann Method (LBM)~\citep{kruger2017lattice}.
LBM-family solvers are inherently suited for unsteady fluid problems and are therefore appropriate for the problems addressed here.
Design sensitivities are computed using the adjoint variable method formulated for LKS, known as the Adjoint Lattice Kinetic Scheme (ALKS)~\citep{TANABE2025114001}.
Because the original ALKS is derived in a continuous formulation in both time and space for stationary, non-moving structures and therefore considers only the fluid-flow equations, this study presents adjoint equations and design sensitivities that couple the fluid-flow equations with the rigid-body equation of motion.
The design sensitivities obtained with the proposed method are also verified by comparison with finite difference approximations.
After obtaining the design sensitivities, the shape is updated using the Method of Moving Asymptotes (MMA)~\citep{svanberg1987method}, a widely used gradient-based optimization solver in topology optimization.
A density filter with Heaviside projection~\citep{wang2011projection} is additionally employed.

This paper is organized as follows:
Section~\ref{sec2} introduces the fundamental equations of the proposed method.
Section~\ref{sec3} explains the details of the numerical implementation.
Section~\ref{sec4} presents numerical examples for two- and three-dimensional problems.
Finally, Section~\ref{sec5} concludes the paper by summarizing the key findings. 

\section{Formulation}
\label{sec2}

\subsection{Topology optimization for fluid and rigid body}
\label{sec21}

\begin{figure}[t]
    \centering
    \includegraphics[width=0.4\columnwidth]{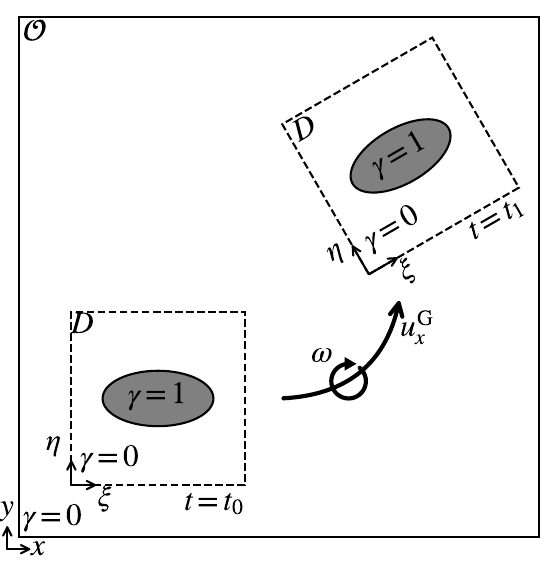}
    \caption{Relationship between the design and analysis domains}
    \label{fig:schematic}
\end{figure}

Topology optimization is a structural optimization method that has attracted considerable attention because it enables substantial changes in structural layout during the design process.
In topology optimization, a structural optimization problem is reformulated as a material distribution problem; that is, the method determines, at each point within a fixed design domain, whether solid or fluid should be placed to minimize or maximize the objective function, subject to constraints.
Here, we optimize the rigid-body shape driven by fluid-flow forces. The design domain $D$ then undergoes rigid-body motion within the analysis domain $\mathcal{O}$, as depicted in Fig.~\ref{fig:schematic}.
The optimization problem is defined as follows:
\begin{align}
    \begin{array}{ll}
        \underset{\gamma}{\text{maximize }} & J\left(\gamma,U\left(\gamma\right)\right)                    \\
        \text{subject to}                   & G\left(\gamma,U\left(\gamma\right)\right)\leq0\vspace{1.5mm} \\
                                            & 0\leq\gamma\left(\bm{\xi}\right)\leq1,
    \end{array} \label{eq:optimization_formulation}
\end{align}
where $\gamma$ and $U$ are the design variable and the state fields.
$\gamma$ is a function of $\bm{\xi}$, the local coordinate in $D$, and takes continuous values between $0$ and $1$.
In the fluid region, $\gamma = 0$, whereas in the solid region, $\gamma = 1$; regions with intermediate values correspond to porous material.
This approach is called the pseudo-density method~\citep{bendsoe2003topology}.
Generally, in the formulation of the optimization problem defined in Eq.~\eqref{eq:optimization_formulation}, multiple constraint functionals are applied; however, in this study, a single constraint, $G$, is applied, and its specific form is described in Section~\ref{sec26}.
It is noted that, in topology optimization for stationary components, the material distribution in the analysis-domain coordinates does not change over time.
However, in the case discussed in this study, the material distribution in the analysis-domain coordinates changes over time, although it does not change in the design-domain coordinates.

The objective functional $J$ and the constraint functional $G$ depend on $\gamma$ and $U$, and, in this study, $U$ consists of the fluid pressure $p$, the fluid velocity $u_\alpha$, the center of gravity of the rigid body $x^\text{G}_\alpha$, the velocity of the center of gravity $u^\text{G}_\alpha$, the rotation angle of the rigid body $\theta$, and the rotational velocity of the rigid body $\omega$.
Here, the subscript $\alpha$ denotes $x$ and $y$ in the two-dimensional case and $x$, $y$, and $z$ in the three-dimensional case.
For simplicity, $\theta$ and $\omega$ are defined with respect to an axis parallel to the $z$-axis; however, the formulation can be easily extended to arbitrary rotations using Euler angles or quaternions.
$p$ and $u_\alpha$ are governed by the continuity and momentum conservation equations as follows:
\begin{align}
     & \frac{\partial u_\alpha}{\partial x_\alpha}=0,                                                                                                                                                                                                                 \\
     & \frac{\partial u_\alpha}{\partial t}+u_\beta\frac{\partial u_\alpha}{\partial x_\beta}=-\frac{\partial p}{\partial x_\alpha}+\nu\frac{\partial^2 u_\alpha}{\partial x_\beta^2}-\kappa\left(u_\alpha-u_\alpha^\text{S}\right), \label{eq:momentum_conservation}
\end{align}
where $\nu$ is the kinematic viscosity of the fluid, and subscripts $\alpha$ and $\beta$ denote $x$ and $y$ in the two-dimensional case and $x$, $y$, and $z$ in the three-dimensional case.
The third term on the right hand side of Eq.~\eqref{eq:momentum_conservation} is called the Brinkman force.
Here, $u^\text{S}_\alpha$ is the velocity of the solid, and $\kappa$ takes a large value only in solid region; therefore, only in the solid region, $u_\alpha$ approaches $u^\text{S}_\alpha$.
The specific forms of $\kappa$ and $u^\text{S}_\alpha$ are stated in Section~\ref{sec23}.

In contrast, $x^\text{G}_\alpha$, $u^\text{G}_\alpha$, $\theta$, and $\omega$ are governed by the equations of motion defined as follows:
\begin{align}
     & \frac{dx^\text{G}_\alpha}{dt}=u^\text{G}_\alpha, \label{eq:equation_of_motion_ug}                                                                                                                                             \\
     & M\frac{du^\text{G}_\alpha}{dt}=\int_D\kappa^\text{ref}\left(u^\text{ref}_\alpha-u^\text{S,ref}_\alpha\right)d\Omega^\text{ref}, \label{eq:equation_of_motion_translation}                                                     \\
     & \frac{d\theta}{dt}=\omega, \label{eq:equation_of_motion_omega}                                                                                                                                                                \\
     & I_z\frac{d\omega}{dt}=\int_De_{z\alpha\beta}\left(x^\text{ref}_\alpha-x^\text{G}_\alpha\right)\kappa^\text{ref}\left(u^\text{ref}_\beta-u^\text{S,ref}_\beta\right)d\Omega^\text{ref}, \label{eq:equation_of_motion_rotation}
\end{align}
where $M$ and $I_z$ are the mass of the rigid body and the moment of inertia about the $z$-axis, respectively, and for simplicity, both are assumed to be constant throughout the optimization process.
$\kappa^\text{ref}$ is the Brinkman coefficient and is related to $\gamma$ as follows:
\begin{equation}
    \kappa^\text{ref}\left(\gamma\right)=\kappa^\text{ref}_\text{max}\frac{q\gamma}{\left(1-\gamma\right)+q}, \label{eq:brinkman_coefficient}
\end{equation}
where $\kappa^\text{ref}_\text{max}$ and $q$ are the maximum value of the Brinkman coefficient and the convexity parameter, respectively.
The right-hand side of Eq.~\eqref{eq:equation_of_motion_translation} and that of Eq.~\eqref{eq:equation_of_motion_rotation} are the force and torque from the fluid flow, defined as the reaction force of Brinkman model in Eq.~\eqref{eq:momentum_conservation}.
The superscript ``ref'' represents quantities at each point in the design domain, and the specific relationships with those without the superscript are described in Section~\ref{sec23}.
$e_{z\alpha\beta}$ is the Eddington $e$ notation, and $x^\text{ref}_\alpha$ represents the analysis-domain coordinates of each point in the design domain.

\subsection{Lattice Kinetic Scheme}
\label{sec22}

\begin{figure}[t]
    \centering
    \begin{minipage}[t]{0.49\columnwidth}
        \centering
        \includegraphics[width=\columnwidth]{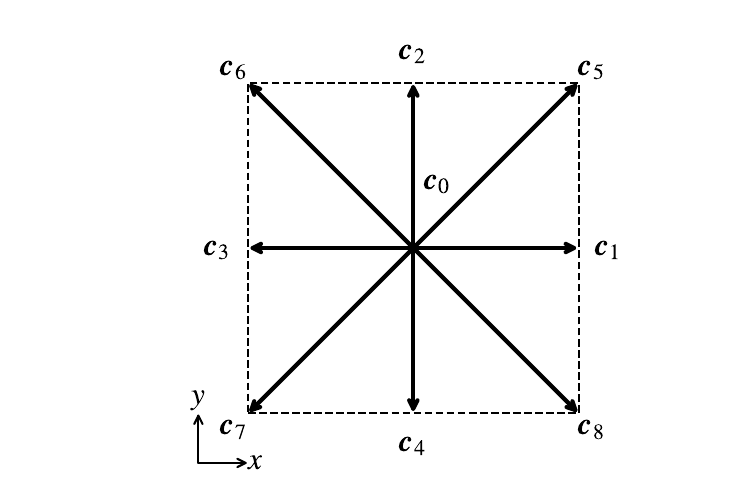}
        \subcaption{D2Q9}
        \label{fig:d2q9}
    \end{minipage}
    \begin{minipage}[t]{0.49\columnwidth}
        \centering
        \includegraphics[width=\columnwidth]{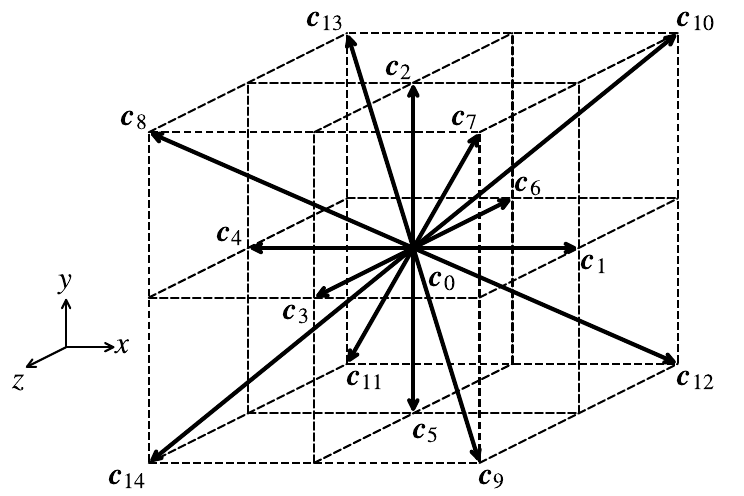}
        \subcaption{D3Q15}
        \label{fig:d3q15}
    \end{minipage}
    \caption{Velocity sets for each lattice gas model.
        Reproduced from Ref.~\citep{TANABE2025114620} under the CC-BY license.}
\end{figure}

The fluid flow is computed using the Lattice Kinetic Scheme (LKS)~\citep{Coveney2002lattice}, which is an extended version of the Lattice Boltzmann Method (LBM)~\citep{kruger2017lattice}.
The basic concept of the LBM-family methods is that fluid flow is approximated by fictitious particles that has a finite set of velocities.
Let $f_i\ \left(i=0,\ldots,Q-1\right)$ be the distribution function of the particles; then they are governed by the discrete-velocity Boltzmann equation described as follows:
\begin{equation}
    \text{Sh}\frac{\partial f_i}{\partial t}+c_{i\alpha}\frac{\partial f_i}{\partial x_\alpha}=-\frac{1}{\varepsilon}\left(f_i-f_i^\text{eq}\right)-3\Delta xw_ic_{i_\alpha}\kappa\left(u_\alpha-u^\text{S}_\alpha\right), \label{eq:discrete_velocity_boltzmann}
\end{equation}
where $\text{Sh}$ and $\varepsilon$ denote the Strouhal number and the dimensionless parameter of the same order as the Knudsen number, respectively.
$Q$ represents the number of finite velocity sets and depends on the lattice gas model.
In this study, the D2Q9 model (see Fig.~\ref{fig:d2q9}) is used for the two-dimensional case, while the D3Q15 model (see Fig.~\ref{fig:d3q15}) is used for the three-dimensional case.
$w_i$ is the weight factor corresponding to the particle with the $i$-th velocity and also depends on the lattice gas model, such as
\begin{equation}
    w_i=\left\{
    \begin{array}{cl}
        \frac{4}{9}  & i=0       \\
        \frac{1}{9}  & i=1,2,3,4 \\
        \frac{1}{36} & i=5,6,7,8
    \end{array}
    \right.,
\end{equation}
for the D2Q9 model, and
\begin{equation}
    w_i=\left\{
    \begin{array}{cl}
        \frac{2}{9}  & i=0           \\
        \frac{1}{9}  & i=1,\ldots,6  \\
        \frac{1}{72} & i=7,\ldots,14
    \end{array}
    \right.,
\end{equation}
for the D3Q15 model.
$f^\text{eq}_i$ is the local equilibrium distribution function, and its specific form is described later.

Equation~\eqref{eq:discrete_velocity_boltzmann} is discretized as follows:
\begin{align}
     & f_i^*\left(\bm{x}+\bm{c}_i\Delta x,t+\Delta t\right)=f_i\left(\bm{x},t\right)-\frac{1}{\tau}\left\{f_i\left(\bm{x},t\right)-f_i^\text{eq}\left(\bm{x},t\right)\right\}, \label{eq:lattice_boltzmann}                                                                    \\
     & f_i\left(\bm{x},t+\Delta t\right)=f_i^*\left(\bm{x},t+\Delta t\right)-3\Delta xw_ic_{i\alpha}\kappa\left(\bm{x},t+\Delta t\right)\left\{u_\alpha\left(\bm{x},t+\Delta t\right)-u_\alpha^\text{S}\left(\bm{x},t+\Delta t\right)\right\}, \label{eq:lattice_boltzmann_ex}
\end{align}
where $\Delta x$ and $\Delta t$ are the grid spacing and time step, respectively.
$\tau$ is the dimensionless relaxation time related to $\varepsilon$ and is given by $\tau=\varepsilon/\Delta x$.
In the LKS, $\tau$ is set to $1$; therefore, Eq.~\eqref{eq:lattice_boltzmann} takes the simple form
\begin{equation}
    f_i^*\left(\bm{x},t\right)=f_i^\text{eq}\left(\bm{x}-\bm{c}_i\Delta x,t-\Delta t\right).
\end{equation}
In this study, $f_i^\text{eq}$ based on the work of Inamuro~\citep{Coveney2002lattice} is used and is defined as follows:
\begin{equation}
    f_i^\text{eq}=w_i\left(3p+3c_{i\alpha}u_\alpha+\frac{9}{2}c_{i_\alpha}u_\alpha c_{i\beta}u_\beta+\frac{3}{2}u_\alpha^2\right)+w_i\Delta xA\left(\frac{\partial u_\beta}{\partial x_\alpha}+\frac{\partial u_\alpha}{\partial x_\beta}\right)c_{i\alpha}c_{i\beta}. \label{eq:local_equilibrium_function}
\end{equation}
Here, $A$ is a parameter related to the kinematic viscosity $\nu$, which is given by
\begin{equation}
    \nu=\left(\frac{1}{6}-\frac{2}{9}A\right)\Delta x.
\end{equation}
The macroscopic values are computed as moments of $f_i$ as follows:
\begin{align}
     & p^*\left(\bm{x},t\right)=\frac{1}{3}\sum_{i=0}^{Q-1}f_i^\text{eq}\left(\bm{x}-\bm{c}_i\Delta x,t-\Delta t\right), \label{eq:p_tmp}        \\
     & u_\alpha^*\left(\bm{x},t\right)=\sum_{i=0}^{Q-1}c_{i\alpha}f_i^\text{eq}\left(\bm{x}-\bm{c}_i\Delta x,t-\Delta t\right). \label{eq:u_tmp}
\end{align}
Since $p^*$ and $u^*_\alpha$ do not include the effect of the external force term, they are updated using the moment of Eq.~\eqref{eq:lattice_boltzmann_ex} as follows:
\begin{align}
     & p\left(\bm{x},t\right)=p^*\left(\bm{x},t\right), \label{eq:p}                                                                                                                                       \\
     & u_\alpha\left(\bm{x},t\right)=\frac{u^*_\alpha\left(\bm{x},t\right)+\Delta x\kappa\left(\bm{x},t\right)u^\text{S}_\alpha\left(\bm{x},t\right)}{1+\Delta x\kappa\left(\bm{x},t\right)}, \label{eq:u}
\end{align}
In summary, in the LKS, the state fields $p$ and $u_\alpha$ are obtained by repeatedly evaluating Eqs.~\eqref{eq:local_equilibrium_function}--\eqref{eq:u} in a time-marching manner.
As can be seen in Eqs.~\eqref{eq:local_equilibrium_function}--\eqref{eq:u}, in the LKS, only the macroscopic values are required, whereas in the LBM, the distribution functions are also required; therefore, the macroscopic values are prescribed directly at the boundaries.

\subsection{Grid separation approach}
\label{sec23}

To represent rigid-body motion in the fluid flow, we use the grid separation approach~\citep{TANABE2025114620}.
The concept of this approach is to separate the design grid, on which the design variables are defined and the design sensitivities are computed, from the analysis grid, on which the state fields are computed.
At each time step, the design grid undergoes rigid-body motion and is then overlapped onto the analysis grid.
To compute the fluid flow, quantities such as the Brinkman coefficient $\kappa$ and the solid velocity at each point in the design domain, $u^\text{S}_\alpha$, are transferred from the design grid to the analysis grid by the equations as follows:
\begin{align}
     & \kappa\left(\bm{x},t\right)=\int_DW\left(\bm{x},\bm{x}^\text{ref}\left(\bm{\xi},t\right)\right)\kappa^\text{ref}\left(\bm{\xi}\right)d\Omega^\text{ref}, \label{eq:kappa_mapping}               \\
     & u^\text{S}_\alpha\left(\bm{x},t\right)=\int_DW\left(\bm{x},\bm{x}^\text{ref}\left(\bm{\xi},t\right)\right)u^\text{S,ref}_\alpha\left(\bm{\xi},t\right)d\Omega^\text{ref}, \label{eq:us_mapping}
\end{align}
where $W$ is a weight function and is defined as follows:
\begin{align}
    W\left(\bm{x},\bm{x}^{\text{ref}}\right) & =\prod_{\alpha=0}^{d-1}w\left(x_\alpha-x_\alpha^{\text{ref}}\right),                                                                                                                              \\
    w\left(r\right)                          & =\left\{\begin{array}{lc}\frac{1}{4\Delta x}\left\{1-\cos{\left(\frac{\pi r}{2}\right)}\right\}&\left|r\right|\leq2\Delta x \\ 0&2\Delta x<\left|r\right|\end{array}\right.,\label{eq:weight_function}
\end{align}
where $d$ denotes the spatial dimension, such as $2$ for the two-dimensional case and $3$ for the three-dimensional case.
Quantities with the superscript ``ref'' are the functions of the local coordinates of the design domain, $\bm{\xi}=\left(\xi,\eta\right)$ for the two-dimensional case and $\bm{\xi}=\left(\xi,\eta,\zeta\right)$ for the three-dimensional case, whereas those without superscript are for the functions of the coordinates in the analysis domain, $\bm{x}$.
In contrast, to compute the motion of the rigid body, fluid velocities are transferred from the analysis grid to the design grid in the same manner as above, as follows:
\begin{equation}
    u^\text{ref}_\alpha\left(\bm{\xi},t\right)=\int_\mathcal{O}W\left(\bm{x},\bm{x}^\text{ref}\left(\bm{\xi},t\right)\right)u_\alpha\left(\bm{x},t\right)d\Omega. \label{eq:u_mapping}
\end{equation}

As mentioned above, the design domain undergoes rigid-body motion; therefore, the design grid, which discretizes the design domain, also undergoes rigid-body motion.
For simplicity, the explanation for the rigid-body motion below is limited to the two-dimensional case; however, the motion can be easily extended to the three-dimensional case.
At each time step, the coordinates of the each point in the design grid in the analysis-domain coordinates are as follows:
\begin{equation}
    \bm{x}^{\text{ref}}\left(\bm{\xi},t\right)=  \left(\begin{array}{cc}
            \cos{\theta\left(t\right)} & -\sin{\theta\left(t\right)} \\
            \sin{\theta\left(t\right)} & \cos{\theta\left(t\right)}
        \end{array}\right)\left(\begin{array}{c}\xi-\xi^{\text{G}}\\\eta-\eta^{\text{G}}\end{array}\right)+\left(\begin{array}{c}x_x^\text{G}\left(t\right)\\x_y^\text{G}\left(t\right)\end{array}\right), \label{eq:rigid_body_position}
\end{equation}
where $\xi^\text{G}$ and $\eta^\text{G}$ are the coordinates of the center point of the rigid-body rotation.
The velocities of each grid point on the design grid are obtained by differentiating Eq.~\eqref{eq:rigid_body_position} with respect to $t$ as follows:
\begin{equation}
    \bm{u}^\text{S,ref}\left(\bm{\xi},t\right)=\left(\begin{array}{cc}
            -\sin{\theta\left(t\right)} & -\cos{\theta\left(t\right)} \\
            \cos{\theta\left(t\right)}  & -\sin{\theta\left(t\right)}
        \end{array}\right)\left(\begin{array}{c}\xi-\xi^{\text{G}}\\\eta-\eta^{\text{G}}\end{array}\right)\omega\left(t\right)+\left(\begin{array}{c}u^\text{G}_x\left(t\right)\\u^\text{G}_y\left(t\right)\end{array}\right). \label{eq:rigid_body_velocity}
\end{equation}
As mentioned in Sec.~\ref{sec21}, in this study the rigid-body motion is governed by the equation of motion; therefore, $x_x^\text{G}$, $x_y^\text{G}$, $u^\text{G}_x$, $u^\text{G}_y$, $\theta$ and $\omega$ are computed by solving Eqs.~\eqref{eq:equation_of_motion_ug}--\eqref{eq:equation_of_motion_rotation} in time, and their discretized forms are described in Section~\ref{sec24}.

\subsection{Equations of motion discretization}
\label{sec24}

The equations of motion in Eqs.~\eqref{eq:equation_of_motion_ug}--\eqref{eq:equation_of_motion_rotation} are discretized using the Euler method as follows:
\begin{align}
     & x^\text{G}_\alpha\left(t\right)=x^\text{G}_\alpha\left(t-\Delta t\right)+u^\text{G}_\alpha\left(t\right), \label{eq:equation_of_motion_xg_discretized}                                                                                                                                                                                                                                  \\
     & u^\text{G}_\alpha\left(t\right)=u^\text{G}_\alpha\left(t-\Delta t\right)+\frac{1}{M}\sum_{m=1}^{N_\text{grid}^\text{ref}}\kappa^\text{ref}\left(\bm{\xi}_m\right)\left\{u^\text{ref}_\alpha\left(\bm{\xi}_m,t\right)-u^\text{S,ref}_\alpha\left(\bm{\xi}_m,t\right)\right\},                                                                                                            \\
     & \theta\left(t\right)=\theta\left(t-\Delta t\right)+\omega\left(t\right),                                                                                                                                                                                                                                                                                                                \\
     & \begin{aligned}
           \omega\left(t\right)= & \omega\left(t-\Delta t\right)                                                                                                                                                                                                                                                                                                                                   \\
                                 & +\frac{1}{I_z}\sum_{m=1}^{N_\text{grid}^\text{ref}}e_{z\alpha\beta}\left\{x^\text{ref}_\alpha\left(\bm{\xi}_m,t\right)-x^\text{G}_\alpha\left(t\right)\right\}\kappa^\text{ref}\left(\bm{\xi}_m\right)\left\{u^\text{ref}_\beta\left(\bm{\xi}_m,t\right)-u^\text{S,ref}_\beta\left(\bm{\xi}_m,t\right)\right\}, \label{eq:equation_of_motion_omega_discretized}
       \end{aligned}
\end{align}
where $N_\text{grid}^\text{ref}$ denotes the number of the grid points on the design grid, and the subscript $m$ indicates quantities defined at the $m$-th grid point.
As previously, subscripts $\alpha$ and $\beta$ denote $x$ and $y$ in the two-dimensional case and $x$, $y$, and $z$ in the three-dimensional case.

\subsection{Sensitivity analysis}
\label{sec25}

In this section, we explain the sensitivity analysis of a general functional $J$ using the ALKS~\citep{TANABE2025114001} coupled with the equations of motion, Eqs.~\eqref{eq:equation_of_motion_ug}--\eqref{eq:equation_of_motion_rotation}.
Here, $J$ is defined as follows:
\begin{equation}
    J=\int_\mathcal{I}\int_\mathcal{O}J_\mathcal{O}d\Omega dt+\int_\mathcal{I}\int_{\partial\mathcal{O}}J_{\partial\mathcal{O}}d\Gamma dt=\int_\mathcal{I}J_\mathcal{I}dt,
\end{equation}
where $J_\mathcal{O}$ and $J_{\partial\mathcal{O}}$ are the integrands defined over the domain and its boundary, respectively, and $J_\mathcal{I}$ is the integrand over the time interval.
First, we define the Lagrangian $L$ as
\begin{align}
    L=   & J +R_1+R_2+R_3+R_4+R_5, \label{eq:lagrangian}                                                                                                                                                                                                                                                            \\
    R_1= & \int_\mathcal{I}\int_\mathcal{O}\sum_{i=0}^{Q-1}\tilde{f}_i\left\{\text{Sh}\frac{\partial f_i}{\partial t}+c_{i\alpha}\frac{\partial f_i}{\partial x_\alpha}+\frac{1}{\varepsilon}\left(f_i-f_i^\text{eq}\right)+3\Delta xw_ic_{i_\alpha}\kappa\left(u_\alpha-u^\text{S}_\alpha\right)\right\}d\Omega dt \\
    R_2= & \int_\mathcal{I}\tilde{x}^\text{G}_\alpha\left(\frac{d x^\text{G}_\alpha}{dt}-u^\text{G}_\alpha\right)dt,                                                                                                                                                                                                \\
    R_3= & \int_\mathcal{I}\tilde{u}^\text{G}_\alpha\left\{M\frac{du^\text{G}_\alpha}{dt}-\int_D\kappa^\text{ref}\left(u^\text{ref}_\alpha-u^\text{S,ref}_\alpha\right)d\Omega^\text{ref}\right\}dt,                                                                                                                \\
    R_4= & \int_\mathcal{I}\tilde{\theta}\left(\frac{d\theta}{dt}-\omega\right)dt,                                                                                                                                                                                                                                  \\
    R_5= & \int_\mathcal{I}\tilde{\omega}\left\{I_z\frac{d\omega}{dt}-\int_De_{z\alpha\beta}\left(x^\text{ref}_\alpha-x^\text{G}_\alpha\right)\kappa^\text{ref}\left(u^\text{ref}_\beta-u^\text{S,ref}_\beta\right)d\Omega^\text{ref}\right\}dt.
\end{align}
Here, $\tilde{f}_i$, $\tilde{x}^\text{G}_\alpha$, $\tilde{u}^\text{G}_\alpha$, $\tilde{\theta}$ and $\tilde{\omega}$ are the Lagrange multipliers.
Strictly speaking, the residuals of the boundary conditions should be included in Eq.~\eqref{eq:lagrangian}; however, for simplicity, these terms are neglected here.

As described in Appendix~\ref{appendixA}, the functional derivative of $L$ with respect to the design variable $\gamma$ is rearranged to avoid explicitly computing the derivatives of the state fields with respect to the design variable.
Consequently, the derivative of $J$, $J^\prime=L^\prime$, is obtained as follows:
\begin{align}
    \langle J^\prime,\delta\gamma\rangle= & \int_\mathcal{I}\int_\mathcal{O}\frac{\partial J_\mathcal{O}}{\partial\gamma}\delta\gamma d\Omega dt+\int_\mathcal{I}\int_{\partial\mathcal{O}}\frac{\partial J_{\partial\mathcal{O}}}{\partial\gamma}\delta\gamma d\Gamma dt \notag                                       \\
                                          & +\int_\mathcal{I}\int_\mathcal{O}\int_D3\Delta x\frac{\partial\kappa^\text{ref}}{\partial\gamma}W\left(u_\alpha-u^\text{S}_\alpha\right)\tilde{u}_\alpha\delta\gamma d\Omega^\text{ref}d\Omega dt\notag                                                                    \\
                                          & -\int_\mathcal{I}\tilde{u}^\text{G}_\alpha\int_D\frac{\partial\kappa^\text{ref}}{\partial\gamma}\delta\gamma\left(u^\text{ref}_\alpha-u^\text{S,ref}_\alpha\right)d\Omega^\text{ref}dt \notag                                                                              \\
                                          & -\int_\mathcal{I}\tilde{\omega}\int_De_{z\alpha\beta}\left(x^\text{ref}_\alpha-x^\text{G}_\alpha\right)\frac{\partial\kappa^\text{ref}}{\partial\gamma}\delta\gamma\left(u^\text{ref}_\beta-u^\text{S,ref}_\beta\right)d\Omega^\text{ref}dt. \label{eq:design_sensitivity}
\end{align}
Here, tilde quantities such as $\tilde{u}_\alpha$ are obtained by solving the adjoint equations, which are derived from the coefficients of $\delta f_i$, $\delta x^\text{G}_\alpha$, $\delta u^\text{G}_\alpha$, $\delta\theta$ and $\delta\omega$, and are discretized in the same manner as the state fields.

The equations used to compute $\tilde{p}$, $\tilde{u}_\alpha$, and $\tilde{s}_{\alpha\beta}$ are given as follows:
\begin{align}
     & \tilde{p}^*\left(\bm{x},t\right)=\sum_{i=0}^{Q-1}w_i\tilde{f}^\text{eq}_i\left(\bm{x}+\bm{c}_i\Delta x,t+\Delta t\right), \label{eq:ap_tmp}                                    \\
     & \tilde{u}^*_\alpha\left(\bm{x},t\right)=\sum_{i=0}^{Q-1}w_ic_{i\alpha}\tilde{f}^\text{eq}_i\left(\bm{x}+\bm{c}_i\Delta x,t+\Delta t\right),                                    \\
     & \tilde{s}^*_{\alpha\beta}\left(\bm{x},t\right)=\sum_{i=0}^{Q-1}w_ic_{i\alpha}c_{i\beta}\tilde{f}^\text{eq}_i\left(\bm{x}+\bm{c}_i\Delta x,t+\Delta t\right), \label{eq:as_tmp}
\end{align}
where $\tilde{f}^\text{eq}_i$ is defined as follows:
\begin{equation}
    \tilde{f}^\text{eq}_i=\tilde{p}+3c_{i\alpha}\left(\tilde{u}_\alpha+3\tilde{s}_{\alpha\beta}u_\beta-\tilde{p}u_\alpha\right)-\Delta xA\frac{\partial}{\partial x_\beta}\left(\tilde{s}_{\alpha\beta}+\tilde{s}_{\beta\alpha}\right)c_{i\alpha}.
\end{equation}
From the quantities with superscript ``*'', $\tilde{p}$, $\tilde{u}_\alpha$, and $\tilde{s}_{\alpha\beta}$ are computed as follows:
\begin{align}
    \tilde{p}\left(\bm{x},t\right)=               & \tilde{p}^*\left(\bm{x},t\right)-\sum_{i=0}^{Q-1}w_i\frac{\partial J_\mathcal{O}}{\partial f_i}\left(\bm{x},t\right),  \label{eq:ap}                                                                                                                                                               \\
    \tilde{u}_\alpha\left(\bm{x},t\right)=        & \frac{1}{1+\Delta x\kappa\left(\bm{x},t\right)}\left\{u^*_\alpha\left(\bm{x},t\right)+\frac{1}{3}\tilde{u}^\text{G}_\alpha\left(t\right)\sum_{m=1}^{N_\text{grid}^\text{ref}}\kappa^\text{ref}\left(\bm{\xi}_m\right)W\left(\bm{x},\bm{x}^\text{ref}\left(\bm{\xi}_m,t\right)\right)\right. \notag \\
                                                  & +\frac{1}{3}\tilde{\omega}\left(t\right)\sum_{m=1}^{N_\text{grid}^\text{ref}}e_{z\beta\alpha}\left(x^\text{ref}_\beta\left(\bm{\xi}_m,t\right)-x^\text{G}_\beta\left(t\right)\right)\kappa^\text{ref}\left(\bm{\xi}_m\right)W\left(\bm{x},\bm{x}^\text{ref}\left(\bm{\xi}_m,t\right)\right) \notag \\
                                                  & \left.-\sum_{i=0}^{Q-1}w_ic_{i\alpha}\frac{\partial J_\mathcal{O}}{\partial f_i}\left(\bm{x},t\right)\right\}, \label{eq:au}                                                                                                                                                                       \\
    \tilde{s}_{\alpha\beta}\left(\bm{x},t\right)= & \tilde{s}^*_{\alpha\beta}\left(\bm{x},t\right)-\sum_{i=0}^{Q-1}w_ic_{i\alpha}c_{i\beta}\frac{\partial J_\mathcal{O}}{\partial f_i}\left(\bm{x},t\right). \label{eq:as}
\end{align}
These equations are referred to as the ALKS.
As in the LKS, only the macroscopic variables are required; therefore, on the boundary, $\tilde{p}$, $\tilde{u}_\alpha$ and $\tilde{s}_{\alpha\beta}$ are prescribed directly.
For simplicity, the detailed description is omitted here.

The equations used to obtain $\tilde{x}^\text{G}_\alpha$, $\tilde{u}^\text{G}_\alpha$, $\tilde{\theta}$, and $\tilde{\omega}$ are given as follows:
\begin{align}
    \tilde{x}^\text{G}_\alpha\left(t\right)= & \tilde{x}^\text{G}_\alpha\left(t+\Delta t\right) \notag                                                                                                                                                                                                                                                                                                                                             \\
                                             & -\sum_{n=1}^{N_\text{grid}}\sum_{m=1}^{N_\text{grid}^\text{ref}}3\Delta x\kappa^\text{ref}\left(\bm{\xi}_m\right)\frac{\partial W}{\partial x^\text{ref}_\alpha}\left(\bm{x}_n,\bm{x}^\text{ref}\left(\bm{\xi}_m,t\right)\right)\left\{u_\beta\left(\bm{x}_n,t\right)-u^\text{S}_\beta\left(\bm{x}_n,t\right)\right\}\tilde{u}_\beta\left(\bm{x}_n,t\right) \notag                                  \\
                                             & +\sum_{n=1}^{N_\text{grid}}\sum_{m=1}^{N_\text{grid}^\text{ref}}3\Delta x\kappa\left(\bm{x}_n,t\right)\tilde{u}_\beta\left(\bm{x}_n,t\right)u^\text{S,ref}_\beta\left(\bm{\xi}_m,t\right)\frac{\partial W}{\partial x^\text{ref}_\alpha}\left(\bm{x}_n,\bm{x}^\text{ref}\left(\bm{\xi}_m,t\right)\right) \notag                                                                                     \\
                                             & +\tilde{u}^\text{G}_\beta\left(t\right)\sum_{m=1}^{N_\text{grid}^\text{ref}}\sum_{n=1}^{N_\text{grid}}\kappa^\text{ref}\left(\bm{\xi}\right)u_\beta\left(\bm{x},t\right)\frac{\partial W}{\partial x^\text{ref}_\alpha}\left(\bm{x}_n,\bm{x}^\text{ref}\left(\bm{\xi}_m,t\right)\right) \notag                                                                                                      \\
                                             & +\tilde{\omega}\left(t\right)\sum_{m=1}^{N_\text{grid}^\text{ref}}\sum_{n=1}^{N_\text{grid}}e_{z\gamma\beta}\left\{x^\text{ref}_\gamma\left(\bm{\xi}_m,t\right)-x^\text{G}_\gamma\left(t\right)\right\}\kappa^\text{ref}\left(\bm{\xi}_m\right)u_\beta\left(\bm{x}_n,t\right)\frac{\partial W}{\partial x^\text{ref}_\alpha}\left(\bm{x}_n,\bm{x}^\text{ref}\left(\bm{\xi}_m,t\right)\right) \notag \\
                                             & -\frac{\partial J_\mathcal{I}}{\partial x^\text{G}_\alpha}\left(t\right), \label{eq:axg}
\end{align}
\begin{align}
    \tilde{u}^\text{G}_\alpha\left(t\right)= & \tilde{u}^\text{G}_\alpha\left(t+\Delta t\right)+\frac{1}{M}\left[\tilde{x}^\text{G}_\alpha\left(t\right)+\sum_{n=1}^{N_\text{grid}}\sum_{m=1}^{N_\text{grid}^\text{ref}}3\Delta x\kappa\left(\bm{x}_n,t\right)\tilde{u}_\alpha\left(\bm{x}_n,t\right)W\left(\bm{x}_n,\bm{x}^\text{ref}\left(\bm{\xi}_m,t\right)\right)\right.\notag                                                                                                          \\
                                             & \left.-\tilde{u}^\text{G}_\alpha\left(t\right)\sum_{m=1}^{N_\text{grid}^\text{ref}}\kappa^\text{ref}\left(\bm{\xi}_m\right)-\tilde{\omega}\left(t\right)\sum_{m=1}^{N_\text{grid}^\text{ref}}e_{z\beta\alpha}\left\{x^\text{ref}_\beta\left(\bm{\xi}_m,t\right)-x^\text{G}_\beta\left(t\right)\right\}\kappa^\text{ref}\left(\bm{\xi}_m\right)-\frac{\partial J_\mathcal{I}}{\partial u^\text{G}_\alpha}\left(t\right)\right], \label{eq:aug}
\end{align}
\begin{align}
    \tilde{\theta}\left(t\right)= & \tilde{\theta}\left(t+\Delta t\right) \notag                                                                                                                                                                                                                                                                                                                                                                                                                                    \\
                                  & -\sum_{n=1}^{N_\text{grid}}\sum_{m=1}^{N_\text{grid}^\text{ref}}3\Delta x\kappa^\text{ref}\left(\bm{\xi}_m\right)\frac{\partial W}{\partial x^\text{ref}_\alpha}\left(\bm{x}_n,\bm{x}^\text{ref}\left(\bm{\xi}_m,t\right)\right)\frac{\partial x^\text{ref}_\alpha}{\partial\theta}\left(\bm{\xi}_m,t\right)\left\{u_\beta\left(\bm{x}_n,t\right)-u^\text{S}_\beta\left(\bm{x}_n,t\right)\right\}\tilde{u}_\beta\left(\bm{x}_n,t\right) \notag                                  \\
                                  & +\sum_{n=1}^{N_\text{grid}}\sum_{m=1}^{N_\text{grid}^\text{ref}}3\Delta x\kappa\left(\bm{x}_n,t\right)\tilde{u}_\alpha\left(\bm{x}_n,t\right)\frac{\partial u^\text{S,ref}_\alpha}{\partial\theta}\left(\bm{\xi}_m,t\right)W\left(\bm{x}_n,\bm{x}^\text{ref}\left(\bm{\xi}_m,t\right)\right) \notag                                                                                                                                                                             \\
                                  & +\sum_{n=1}^{N_\text{grid}}\sum_{m=1}^{N_\text{grid}^\text{ref}}3\Delta x\kappa\left(\bm{x}_n,t\right)\tilde{u}_\alpha\left(\bm{x}_n,t\right)u^\text{S,ref}_\alpha\left(\bm{\xi}_m,t\right)\frac{\partial W}{\partial x^\text{ref}_\beta}\left(\bm{x}_n,\bm{x}^\text{ref}\left(\bm{\xi}_m,t\right)\right)\frac{\partial x^\text{ref}_\beta}{\partial\theta}\left(\bm{\xi}_m,t\right)\notag                                                                                      \\
                                  & +\tilde{u}^\text{G}_\alpha\left(t\right)\sum_{m=1}^{N_\text{grid}^\text{ref}}\sum_{n=1}^{N_\text{grid}}\kappa^\text{ref}\left(\bm{\xi}_m\right)u_\alpha\left(\bm{x}_n,t\right)\frac{\partial W}{\partial x^\text{ref}_\beta}\left(\bm{x}_n,\bm{x}^\text{ref}\left(\bm{\xi}_m,t\right)\right)\frac{\partial x^\text{ref}_\beta}{\partial\theta}\left(\bm{\xi}_m,t\right) \notag                                                                                                  \\
                                  & -\tilde{u}^\text{G}_\alpha\left(t\right)\sum_{m=1}^{N_\text{grid}^\text{ref}}\kappa^\text{ref}\left(\bm{\xi}_m\right)\frac{\partial u^\text{S,ref}_\alpha}{\partial\theta}\left(\bm{\xi}_m,t\right)\notag                                                                                                                                                                                                                                                                       \\
                                  & +\tilde{\omega}\left(t\right)\sum_{m=1}^{N_\text{grid}^\text{ref}}e_{z\alpha\beta}\frac{\partial x^\text{ref}_\alpha}{\partial\theta}\left(\bm{\xi}_m,t\right)\kappa^\text{ref}\left(\bm{\xi}_m\right)\left\{u^\text{ref}_\beta\left(\bm{\xi}_m,t\right)-u^\text{S,ref}_\beta\left(\bm{\xi}_m,t\right)\right\}\notag                                                                                                                                                            \\
                                  & +\tilde{\omega}\left(t\right)\sum_{m=1}^{N_\text{grid}^\text{ref}}\sum_{n=1}^{N_\text{grid}}e_{z\alpha\beta}\left\{x^\text{ref}_\alpha\left(\bm{\xi}_m,t\right)-x^\text{G}_\alpha\left(t\right)\right\}\kappa^\text{ref}\left(\bm{\xi}_m\right)u_\beta\left(\bm{x}_n,t\right)\frac{\partial W}{\partial x^\text{ref}_\gamma}\left(\bm{x}_n,\bm{x}^\text{ref}\left(\bm{\xi}_m,t\right)\right)\frac{\partial x^\text{ref}_\gamma}{\partial\theta}\left(\bm{\xi}_m,t\right) \notag \\
                                  & -\tilde{\omega}\left(t\right)\sum_{m=1}^{N_\text{grid}^\text{ref}}e_{z\alpha\beta}\left\{x^\text{ref}_\alpha\left(\bm{\xi}_m,t\right)-x^\text{G}_\alpha\left(t\right)\right\}\kappa^\text{ref}\left(\bm{\xi}_m\right)\frac{\partial u^\text{S,ref}_\beta}{\partial\theta}\left(\bm{\xi}_m,t\right)-\frac{\partial J_\mathcal{I}}{\partial\theta}\left(t\right), \label{eq:atheta}
\end{align}
\begin{align}
    \tilde{\omega}\left(t\right)= & \tilde{\omega}\left(t+\Delta t\right)+\frac{1}{I_z}\left[\tilde{\theta}\left(t\right)\right.\notag                                                                                                                                                                                                                                                                                             \\
                                  & +\sum_{n=1}^{N_\text{grid}}\sum_{m=1}^{N_\text{grid}^\text{ref}}3\Delta x\kappa\left(\bm{x}_n,t\right)\tilde{u}_\alpha\left(\bm{x}_n,t\right)\frac{\partial u^\text{S,ref}_\alpha}{\partial\omega}\left(\bm{\xi}_m,t\right)W\left(\bm{x}_n,\bm{x}^\text{ref}\left(\bm{\xi}_m,t\right)\right)\notag                                                                                             \\
                                  & -\tilde{u}^\text{G}_\alpha\left(t\right)\sum_{m=1}^{N_\text{grid}^\text{ref}}\kappa^\text{ref}\left(\bm{\xi}_m\right)\frac{\partial u^\text{S,ref}}{\partial\omega}\left(\bm{\xi}_m,t\right) \notag                                                                                                                                                                                            \\
                                  & \left.-\tilde{\omega}\left(t\right)\sum_{m=1}^{N_\text{grid}^\text{ref}}e_{z\alpha\beta}\left\{x^\text{ref}_\alpha\left(\bm{\xi}_m,t\right)-x^\text{G}_\alpha\left(t\right)\right\}\kappa^\text{ref}\left(\bm{\xi}_m\right)\frac{\partial u^\text{S,ref}_\beta}{\partial\omega}\left(\bm{\xi}_m,t\right)-\frac{\partial J_\mathcal{I}}{\partial\omega}\left(t\right)\right], \label{eq:aomega}
\end{align}
where the derivatives of $x^\text{ref}_\alpha$ with respect to $\theta$ are, according to Eq.~\eqref{eq:rigid_body_position}, given as follows:
\begin{equation}
    \left(\begin{array}{c}
        \partial x^\text{ref}_x/\partial\theta \\
        \partial x^\text{ref}_y/\partial\theta
    \end{array}\right)=  \left(\begin{array}{cc}
        -\sin{\theta\left(t\right)} & -\cos{\theta\left(t\right)} \\
        \cos{\theta\left(t\right)}  & -\sin{\theta\left(t\right)}
    \end{array}\right)\left(\begin{array}{c}\xi-\xi^{\text{G}}\\\eta-\eta^{\text{G}}\end{array}\right).
\end{equation}
Similarly, based on Eq.~\eqref{eq:rigid_body_velocity}, the derivatives of $u^\text{S,ref}_\alpha$ with respect to $\theta$ and $\omega$ are given as follows:
\begin{align}
     & \left(\begin{array}{c}
                 \partial u^\text{S,ref}_x/\partial\theta \\
                 \partial u^\text{S,ref}_y/\partial\theta
             \end{array}\right)=\left(\begin{array}{cc}
                                          -\cos{\theta\left(t\right)} & \sin{\theta\left(t\right)}  \\
                                          -\sin{\theta\left(t\right)} & -\cos{\theta\left(t\right)}
                                      \end{array}\right)\left(\begin{array}{c}\xi-\xi^{\text{G}}\\\eta-\eta^{\text{G}}\end{array}\right)\omega\left(t\right), \\
     & \left(\begin{array}{c}
                 \partial u^\text{S,ref}_x/\partial\omega \\
                 \partial u^\text{S,ref}_y/\partial\omega
             \end{array}\right)=\left(\begin{array}{cc}
                                          -\sin{\theta\left(t\right)} & -\cos{\theta\left(t\right)} \\
                                          \cos{\theta\left(t\right)}  & -\sin{\theta\left(t\right)}
                                      \end{array}\right)\left(\begin{array}{c}\xi-\xi^{\text{G}}\\\eta-\eta^{\text{G}}\end{array}\right).
\end{align}

\subsection{Objective functionals and constraint functional}
\label{sec26}

In this study, we consider two types of objective functionals.
One is the translational distance of the rigid-body center of gravity, defined as follows:
\begin{equation}
    J_1=x^\text{G}_\alpha\left(t_1\right)=\int_\mathcal{I}u^\text{G}_\alpha dt. \label{eq:objective1}
\end{equation}
This functional is used in Section~\ref{sec41}.
The other functional is the rotation angle of the rigid body, defined as follows:
\begin{equation}
    J_2=\theta\left(t_1\right)=\int_\mathcal{I}\omega dt. \label{eq:objective2}
\end{equation}
This functional is used in Sections~\ref{sec42} and \ref{sec43}.

In all examples, the minimum volume of the solid is constrained as follows:
\begin{equation}
    G=\frac{\int_D\gamma d\Omega^\text{ref}}{V_\text{min}\int_Dd\Omega^\text{ref}}-1\leq0, \label{eq:constraint}
\end{equation}
where $V_\text{min}$ denotes the minimum volume ratio.
In typical topology optimization problems, the maximum value of the solid is constrained; however, in the present study, the opposite constraint is imposed.
This is because the problems considered in this study may generate thin structural features, such as airfoil trailing edges.
This treatment has also been employed in airfoil optimization studies~\citep{Kondoh2012}. 

\section{Numerical implementation}
\label{sec3}

In this section, we explain the detail for numerical implementation, such as iterative solution of fluid-rigid-body interaction, filtering scheme, continuation scheme, periodic problem approximation, and optimization procedure.

\subsection{Iterative solution of fluid-rigid-body interaction}
\label{sec31}

As shown in Eqs.~\eqref{eq:lattice_boltzmann_ex}, \eqref{eq:equation_of_motion_xg_discretized}--\eqref{eq:equation_of_motion_omega_discretized}, \eqref{eq:au}, and \eqref{eq:axg}--\eqref{eq:aomega}, the interaction between the fluid flow and the rigid-body motion requires the quantities at the next time step to be mutually dependent.
Since it is difficult to solve this relationship analytically, these quantities are computed iteratively for $N_\text{itr}$ iterations.
This treatment has also been employed in the literature~\citep{SUZUKI2011173}.

\subsection{Filtering scheme}
\label{sec32}

We employ the density filter with a Heaviside projection~\citep{wang2011projection} to remove grayscale regions.
First, the design variable $\gamma$ is filtered as follows:
\begin{equation}
    \tilde{\gamma}_i=\frac{\sum_jw\left(\bm{x}_i,\bm{x}_j\right)\gamma_j}{\sum_jw\left(\bm{x}_i,\bm{x}_j\right)}, \label{eq:filter}
\end{equation}
where the subscripts $i$ and $j$ denote quantities defined at the $i$-th and $j$-th grid points, respectively.
The filter weight $w$ is defined as
\begin{equation}
    w\left(\bm{x}_i,\bm{x}_j\right)=\left\{
    \begin{array}{ll}
        \frac{R-\left\|\bm{x}_i-\bm{x}_j\right\|}{R} & \left\|\bm{x}_i-\bm{x}_j\right\|\leq R \\
        0                                            & \left\|\bm{x}_i-\bm{x}_j\right\|> R
    \end{array}
    \right.,
\end{equation}
where $R$ is the filter radius.
Next, the filtered variable is projected as follows:
\begin{equation}
    \bar{\tilde{\gamma}}_i=\frac{\tanh{\left(\beta\eta\right)}+\tanh{\left(\beta\left(\tilde{\gamma}_i-\eta\right)\right)}}{\tanh{\left(\beta\eta\right)}+\tanh{\left(\beta\left(1-\eta\right)\right)}}, \label{eq:projection}
\end{equation}
where $\beta$ and $\eta$ are parameters controlling the steepness and the threshold, respectively.
It is worth noting that $\bar{\tilde{\gamma}}$ is substituted into the Brinkman coefficient defined as Eq.~\eqref{eq:brinkman_coefficient} and the constraint functional defined as Eq.~\eqref{eq:constraint}, whereas $\gamma$ is updated using a gradient-based optimizer.

\subsection{Continuation scheme}
\label{sec33}

In the optimization, we employ a continuation scheme~\citep{Stolpe2001} for the convexity parameter $q$ in Eq.~\eqref{eq:brinkman_coefficient} and the Heaviside projection parameter $\beta$ in Eq.~\eqref{eq:projection}.
In this scheme, the optimization problem is first solved with fixed values of $q^l$ and $\beta^l$, where the superscript $l$ denotes the continuation step index.
After the stopping criterion is satisfied subject to the constraint, or the maximum number of optimization iterations $N_\text{opt}$ is reached, $q$ and $\beta$ are updated to $q^{l+1}$ and $\beta^{l+1}$, respectively, and the optimization problem is solved again with the updated parameters.
This procedure is repeated until $l=N_\text{cont}$.

In this study, the stopping criterion is defined as follows:
\begin{equation}
    \left|\frac{J^k-J^{k-1}}{J^{k-1}}\right|\leq 10^{-6}, \label{eq:stopping_criterion}
\end{equation}
where the superscript $k$ denotes the optimization step.

\subsection{Periodic problem approximation}
\label{sec34}

For periodic problems, the terminal values of the state fields, such as $p$, $u_\alpha$, $x^\text{G}_\alpha$, $u^\text{G}_\alpha$, $\theta$, and $\omega$, as well as the adjoint fields, such as $\tilde{p}$, $\tilde{u}_\alpha$, $\tilde{s}_{\alpha\beta}$, $\tilde{x}^\text{G}_\alpha$, $\tilde{u}^\text{G}_\alpha$, $\tilde{\theta}$, and $\tilde{\omega}$, at the $k$-th optimization iteration are used as the initial values for the $(k+1)$-th iteration, as in a previous study~\citep{Tanabe2023}.
For example, the initial value of $p$, one of the state fields, at the $(k+1)$-th optimization iteration is given by
\begin{equation}
    p^{k+1}\left(t_0\right)=p^k\left(t_1\right),
\end{equation}
where the superscripts $k$ and $k+1$ denote the optimization iterations.
Similarly, the initial value of $\tilde{p}$, one of the adjoint fields, at the $(k+1)$-th optimization iteration is given by
\begin{equation}
    \tilde{p}^{k+1}\left(t_1\right)=\tilde{p}^k\left(t_0\right).
\end{equation}
Strictly speaking, the effect of the initial conditions should be taken into account; however, this effect is expected to be small when the change in the shape between successive iterations is small, that is, when a small move limit is employed in the optimizer.
By using this approximation, the computation cost is reduced; in other words, without this approximation, additional computations for the transient and some cycles of pseudo-steady phases would be required.
This approximation is employed in the numerical examples presented in Sections~\ref{sec42} and \ref{sec43}.

\subsection{Optimization procedure}
\label{sec35}

The optimization procedure is summarized in Algorithm.~\ref{algorithm:optimization_procedure}.
As shown in Algorithm.~\ref{algorithm:optimization_procedure}, the Method of Moving Asymptotes (MMA)~\citep{svanberg1987method} is employed as the gradient-based optimization solver.

\begin{algorithm}
    \small

    \SetKw{KwInit}{Initialize}

    \caption{Optimization procedure}

    \KwInit{$\gamma$}\;
    \For{$l=1$ to $N_\text{cont}$}{
        $\beta=\beta^\text{l}$\;
        \For{$k=1$ to $N_\text{opt}$}{
            \tcp{Filter variables and inject design variable into governing equations}
            $\gamma\mapsto\tilde{\gamma},\tilde{\gamma},\beta\mapsto\bar{\tilde{\gamma}},\bar{\tilde{\gamma}}\mapsto\kappa^\text{ref}$\tcp*{Eqs.~\eqref{eq:filter}, \eqref{eq:projection} and Eq.~\eqref{eq:brinkman_coefficient}}
            \BlankLine
            \tcp{Compute state fields}
            \KwInit{$p\left(t_0\right)$, $u_\alpha\left(t_0\right)$, $x^\text{G}_\alpha\left(t_0\right)$, $u^\text{G}_\alpha\left(t_0\right)$, $\theta\left(t_0\right)$, $\omega\left(t_0\right)$}\;
            \For{$t=1$ to $N_\text{t}$}{
                $p\left(t-\Delta t\right),u_\alpha\left(t-\Delta t\right)\mapsto p^*\left(t\right),u_\alpha^*\left(t\right)$\tcp*{Eqs.~\eqref{eq:p_tmp} and \eqref{eq:u_tmp}}\
                \For{$itr=1$ to $N_\text{itr}$}{
                    \tcp{Rigid body $\rightarrow$ fluid}
                    $x^\text{G}_\alpha\left(t\right),u^\text{G}_\alpha\left(t\right),\theta\left(t\right),\omega\left(t\right)\mapsto x^\text{ref}_\alpha\left(t\right),u^\text{S,ref}_\alpha\left(t\right)$\tcp*{Eqs.~\eqref{eq:rigid_body_position} and \eqref{eq:rigid_body_velocity}}\
                    $\kappa^\text{ref},x^\text{ref}_\alpha\left(t\right),u^\text{S,ref}_\alpha\left(t\right)\mapsto\kappa\left(t\right),u^\text{S}_\alpha\left(t\right)$\tcp*{Eqs.~\eqref{eq:kappa_mapping} and \eqref{eq:us_mapping}}\
                    $p^*\left(t\right),u^*_\alpha\left(t\right),\kappa,u^\text{S}_\alpha\left(t\right)\mapsto p\left(t\right),u_\alpha\left(t\right)$\tcp*{Eqs.~\eqref{eq:p} and \eqref{eq:u}}
                    \BlankLine
                    \tcp{Fluid $\rightarrow$ rigid body}
                    $u_\alpha\left(t\right)\mapsto u^\text{ref}_\alpha\left(t\right)$\tcp*{Eq.~\eqref{eq:u_mapping}}\
                    $u^\text{ref}_\alpha\left(t\right)\mapsto x^\text{G}_\alpha\left(t\right),u^\text{G}_\alpha\left(t\right),\theta\left(t\right),\omega\left(t\right)$\tcp*{Eqs.~\eqref{eq:equation_of_motion_xg_discretized}-\eqref{eq:equation_of_motion_omega_discretized}}
                }
            }
            \BlankLine
            \tcp{Compute objective and constraint and converse check}
            $p,u_\alpha,x^\text{G}_\alpha,u^\text{G}_\alpha,\theta,\omega\mapsto J,G$\;
            \If{Is stopping criterion satisfied}{
                Bleak\;
            }
            \BlankLine
            \tcp{Compute adjoint fields}
            \KwInit{$\tilde{p}\left(t_1\right)$, $\tilde{u}_\alpha\left(t_1\right)$, $\tilde{s}_{\alpha\beta}\left(t_1\right)$, $\tilde{x}^\text{G}_\alpha\left(t_1\right)$, $\tilde{u}^\text{G}_\alpha\left(t_1\right)$, $\tilde{\theta}\left(t_1\right)$, $\tilde{\omega}\left(t_1\right)$}\;
            \For{$t=1$ to $N_\text{t}$}{
                $\tilde{p}\left(t+\Delta t\right),\tilde{u}_\alpha\left(t+\Delta t\right),\tilde{s}_{\alpha\beta}\left(t+\Delta t\right)\mapsto\tilde{p}^*\left(t\right),\tilde{u}^*_\alpha\left(t\right),\tilde{s}^*_{\alpha\beta}\left(t\right)$\tcp*{Eqs.~\eqref{eq:ap_tmp}-\eqref{eq:as_tmp}}\
                $x^\text{G}_\alpha\left(t\right),u^\text{G}_\alpha\left(t\right),\theta\left(t\right),\omega\left(t\right)\mapsto x^\text{ref}_\alpha\left(t\right),u^\text{S,ref}_\alpha\left(t\right)$\tcp*{Eqs.~\eqref{eq:rigid_body_position} and \eqref{eq:rigid_body_velocity}}\
                $\kappa^\text{ref},x^\text{ref}_\alpha\left(t\right),u^\text{S,ref}_\alpha\left(t\right)\mapsto\kappa\left(t\right),u^\text{S}_\alpha\left(t\right)$\tcp*{Eqs.~\eqref{eq:kappa_mapping} and \eqref{eq:us_mapping}}\
                $u_\alpha\left(t\right)\mapsto u^\text{ref}_\alpha\left(t\right)$\tcp*{Eq.~\eqref{eq:u_mapping}}\
                \For{$itr=1$ to $N_\text{itr}$}{
                    \tcp{Rigid body $\rightarrow$ fluid}
                    $\tilde{p}^*\left(t\right),\tilde{u}^*_\alpha\left(t\right),\tilde{s}^*_{\alpha\beta}\left(t\right),\tilde{u}^\text{G}_\alpha\left(t\right),\tilde{\omega}\left(t\right)\mapsto\tilde{p}\left(t\right),\tilde{u}_\alpha\left(t\right),\tilde{s}_{\alpha\beta}\left(t\right)$\tcp*{Eqs.~\eqref{eq:ap}-\eqref{eq:as}}
                    \BlankLine
                    \tcp{Fluid $\rightarrow$ rigid body}
                    $\tilde{p}\left(t\right),\tilde{u}_\alpha\left(t\right),\tilde{s}_{\alpha\beta}\left(t\right)\mapsto\tilde{x}^\text{G}_\alpha\left(t\right),\tilde{u}^\text{G}_\alpha\left(t\right),\tilde{\theta}\left(t\right),\tilde{\omega}\left(t\right)$\tcp*{Eqs.~\eqref{eq:axg}-\eqref{eq:aomega}}
                }
            }
            \BlankLine
            \tcp{Compute design sensitivity}
            $u_\alpha,x^\text{G}_\alpha,u^\text{G}_\alpha,\theta,\omega,\tilde{u}_\alpha,\tilde{u}^\text{G}_\alpha,\tilde{\omega}\mapsto dJ/d\bar{\tilde{\gamma}},dG/d\bar{\tilde{\gamma}}$\tcp*{Eq.~\eqref{eq:design_sensitivity}}
            \BlankLine
            \tcp{Unfilter design sensitivity and update design variables}
            $dJ/d\bar{\tilde{\gamma}}\mapsto dJ/d\gamma,dG/d\bar{\tilde{\gamma}}\mapsto dG/d\gamma,\gamma^k,dJ/d\gamma,G,dG/d\gamma\mapsto\gamma^{k+1}$\tcp*{MMA}
        }
    }

    \label{algorithm:optimization_procedure}
\end{algorithm}

\section{Numerical example}
\label{sec4}

In this section, we present two two-dimensional examples and one three-dimensional example.
All examples are implemented in an in-house CUDA/C++ code and computed on a single NVIDIA GeForce RTX 3050 6GB Laptop GPU for the two-dimensional cases and on a single NVIDIA GeForce RTX 4070 GPU for the three-dimensional case.

\subsection{2D sail}
\label{sec41}

\begin{figure}[t]
    \centering
    \begin{minipage}[t]{0.49\columnwidth}
        \centering
        \includegraphics[width=0.6\columnwidth]{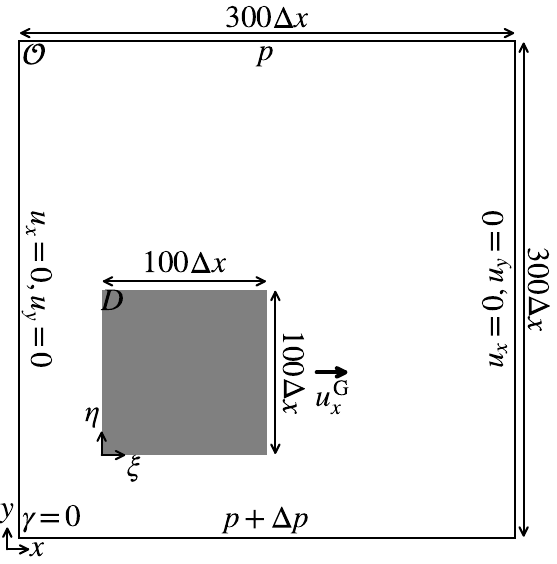}
        \subcaption{Design setting}
        \label{fig:example1_schematic}
    \end{minipage}
    \begin{minipage}[t]{0.49\columnwidth}
        \centering
        \includegraphics[width=0.6\columnwidth]{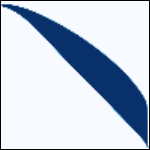}
        \subcaption{Optimized shape}
        \label{fig:example1_optimized_shape}
    \end{minipage}
    \caption{Design setting and optimized shape of the \textit{2D sail}.}
\end{figure}
\begin{figure}[t]
    \centering
    \begin{minipage}[t]{0.49\columnwidth}
        \centering
        \includegraphics[width=0.8\columnwidth]{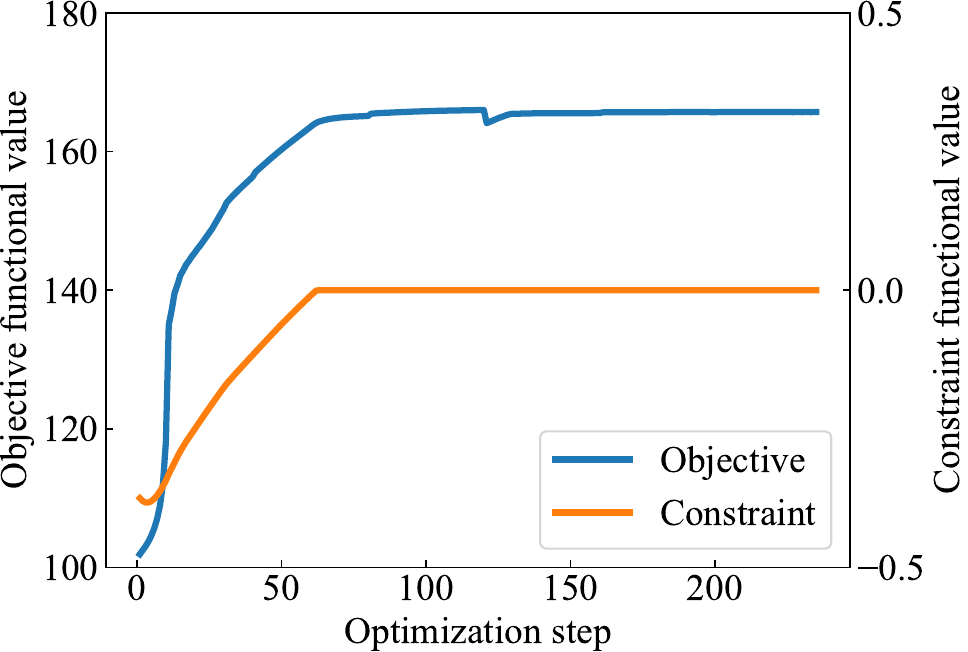}
        \subcaption{Histories of the objective and constraint functional values.}
        \label{fig:example1_history}
    \end{minipage}
    \begin{minipage}[t]{0.49\columnwidth}
        \centering
        \includegraphics[width=0.8\columnwidth]{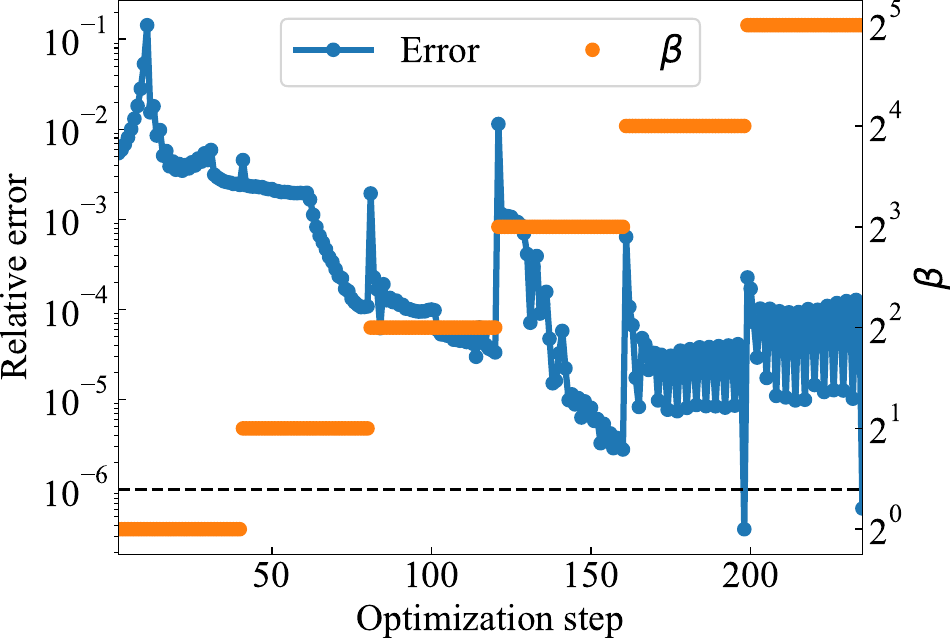}
        \subcaption{Histories of the relative error and steepness parameter.}
        \label{fig:example1_convergence}
    \end{minipage}
    \caption{Histories of the objective and constraint functional values, and the relative error and steepness parameter}
\end{figure}
\begin{figure}[t]
    \centering
    \begin{minipage}[t]{0.255\columnwidth}
        \centering
        \includegraphics[width=\columnwidth]{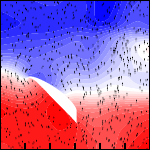}
        \subcaption{$t=1000\Delta t$}
    \end{minipage}
    \begin{minipage}[t]{0.065\columnwidth}
        \centering
        \includegraphics[width=\columnwidth]{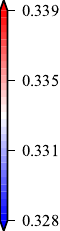}
    \end{minipage}
    \begin{minipage}[t]{0.255\columnwidth}
        \centering
        \includegraphics[width=\columnwidth]{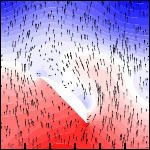}
        \subcaption{$t=2000\Delta t$}
    \end{minipage}
    \begin{minipage}[t]{0.065\columnwidth}
        \centering
        \includegraphics[width=\columnwidth]{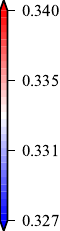}
    \end{minipage}
    \begin{minipage}[t]{0.255\columnwidth}
        \centering
        \includegraphics[width=\columnwidth]{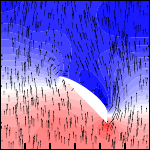}
        \subcaption{$t=3000\Delta t$}
    \end{minipage}
    \begin{minipage}[t]{0.065\columnwidth}
        \centering
        \includegraphics[width=\columnwidth]{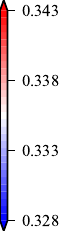}
    \end{minipage}
    \caption{Pressure distribution and flow velocity vector plots for the optimized shape of the \textit{2D sail} at each time step.}
    \label{fig:example1_opt_pressure}
\end{figure}
\begin{figure}[t]
    \centering
    \begin{minipage}[t]{0.33\columnwidth}
        \centering
        \includegraphics[width=0.9\columnwidth]{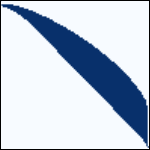}
        \subcaption{Binarized shape}
        \label{fig:example1_binalized_shape}
    \end{minipage}
    \begin{minipage}[t]{0.33\columnwidth}
        \centering
        \includegraphics[width=0.9\columnwidth]{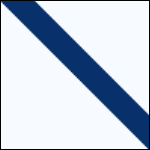}
        \subcaption{Reference shape}
        \label{fig:example1_reference_shape}
    \end{minipage}
    \begin{minipage}[t]{0.325\columnwidth}
        \centering
        \includegraphics[width=\columnwidth]{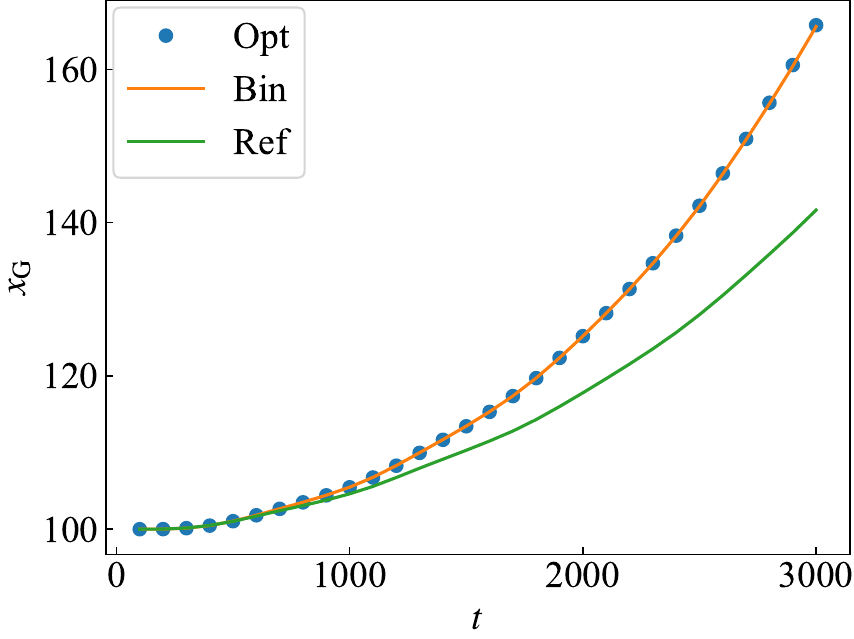}
        \subcaption{Displacement}
        \label{fig:example1_check}
    \end{minipage}
    \caption{Reference shape and binarized shape, and the displacements of the optimized, binarized, and reference shapes for the \textit{2D sail}}
\end{figure}
\begin{figure}[t]
    \centering
    \begin{minipage}[t]{0.255\columnwidth}
        \centering
        \includegraphics[width=\columnwidth]{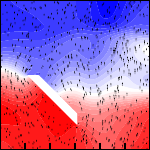}
        \subcaption{$t=1000\Delta t$}
    \end{minipage}
    \begin{minipage}[t]{0.065\columnwidth}
        \centering
        \includegraphics[width=\columnwidth]{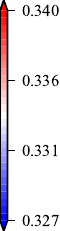}
    \end{minipage}
    \begin{minipage}[t]{0.255\columnwidth}
        \centering
        \includegraphics[width=\columnwidth]{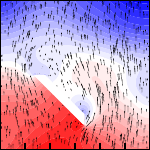}
        \subcaption{$t=2000\Delta t$}
    \end{minipage}
    \begin{minipage}[t]{0.065\columnwidth}
        \centering
        \includegraphics[width=\columnwidth]{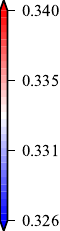}
    \end{minipage}
    \begin{minipage}[t]{0.255\columnwidth}
        \centering
        \includegraphics[width=\columnwidth]{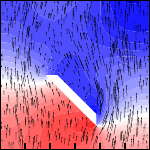}
        \subcaption{$t=3000\Delta t$}
    \end{minipage}
    \begin{minipage}[t]{0.065\columnwidth}
        \centering
        \includegraphics[width=\columnwidth]{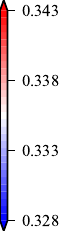}
    \end{minipage}
    \caption{Pressure distribution and flow velocity vector plots for the reference shape of the \textit{2D sail} at each time step.}
    \label{fig:example1_ref_pressure}
\end{figure}

As the first numerical example, we present the optimization of a rigid-body shape driven by fluid flow force, referred to as the \textit{2D sail}.
The design setting is shown in Fig.~\ref{fig:example1_schematic}.
In this example, the rigid body is allowed to move only in the horizontal direction and rotation is prohibited; in other words, only Eqs.~\eqref{eq:equation_of_motion_ug} and \eqref{eq:equation_of_motion_translation} for $\alpha=x$ are considered.
The objective functional is the horizontal displacement given in Eq.~\eqref{eq:objective1} with $\alpha=x$, whereas the constraint is minimum solid volume limitation given in Eq.~\eqref{eq:constraint}.
The parameters are as follows: $N_\text{t}=3000\Delta t$, $N_\text{opt}=40$, $N_\text{itr}=5$, $\nu=0.1$, $M=10^4$, $\Delta p=10^{-2}$, $\kappa^\text{ref}_\text{max}=300$, $q=\left\{0.1,0.1,0.1,1.0,1.0,1.0\right\}$, $\beta=\left\{1,2,4,8,16,32\right\}$ and $V_\text{min}=0.2$.

The optimized shape is shown in Fig.~\ref{fig:example1_optimized_shape}, and the histories of the objective and constraint functional values are shown in Fig.~\ref{fig:example1_history}.
In addition, Fig.~\ref{fig:example1_convergence} shows the relative error, defined as the left-hand side of the stopping criterion in Eq.~\eqref{eq:stopping_criterion}, and the steepness parameter at each optimization step.
The objective functional values, shown in Fig.~\ref{fig:example1_history}, increase almost monotonically while satisfying the constraint and become nearly steady after approximately $60$th optimization step, except for a deviation observed at the $121$st step.
This abrupt change in the objective functional values occurs because the hyperparameters, such as $q$ and $\beta$, are updated, as shown in Fig.~\ref{fig:example1_convergence}.
The computational time per optimization step was approximately $76$~s.

The pressure distributions together with the flow velocity vector plots of the optimized shape at each time step are shown in Fig.~\ref{fig:example1_opt_pressure}.
The optimized shape resembles an airfoil; therefore, the pressure on the upper surface becomes lower, and a lifting force is generated by the fluid flow.
As a result, the rigid body moves horizontally under this force.
It is worth noting that the rigid body accelerates due to the force, and the effective angle of attack changes gradually.
Therefore, unsteady analysis is required to optimize such problems.

In Fig.~\ref{fig:example1_binalized_shape}, the binarized shape is shown, which is obtained by modifying the optimized shape in Fig.~\ref{fig:example1_optimized_shape} using a threshold value of $0.5$.
In addition, a reference shape with almost the same volume as the optimized shape is shown in Fig.~\ref{fig:example1_reference_shape}.
The displacement histories at each time step for each shape are compared in Fig.~\ref{fig:example1_check}.
Here, ``Opt'', ``Bin'' and ``Ref'' denote the displacements of the optimized shape, the binarized shape and the reference shape, respectively.
As shown in Fig.~\ref{fig:example1_check}, there is only a small difference between the displacement of the optimized shape and that of the binarized shape.
This indicates that, although some grayscale regions remain in the optimized shape, their effect is negligible.
On the other hand, a significant difference is observed between the displacement of the optimized and reference shapes.
Fig.~\ref{fig:example1_ref_pressure} presents the pressure distribution together with the flow velocity vector plots for the reference shape at each time step.
As shown in Fig.~\ref{fig:example1_ref_pressure}, the upper surface of the reference shape is flat, resulting in a narrower low-pressure region and a weaker lifting force compared with the optimized shape, whose upper surface is curved.

\subsection{2D turbine}
\label{sec42}

\begin{figure}[t]
    \centering
    \includegraphics[width=0.5\columnwidth]{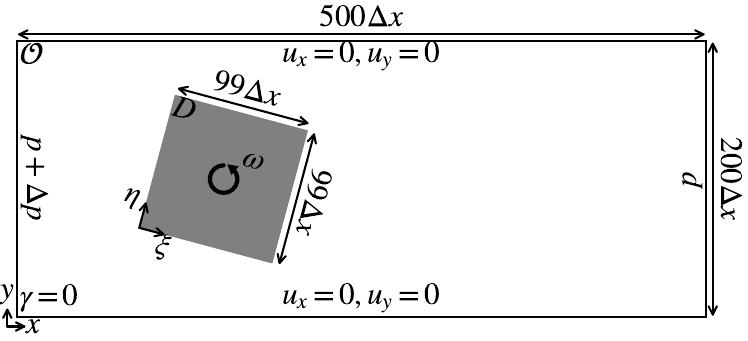}
    \caption{Design setting of the \textit{2D turbine}.}
    \label{fig:example2_schematic}
\end{figure}
\begin{figure}[t]
    \centering
    \includegraphics[width=0.8\columnwidth]{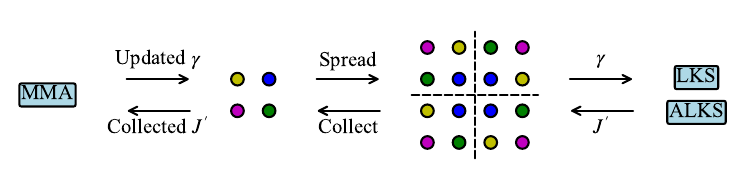}
    \caption{Cyclic periodic mapping of design variable}
    \label{fig:example2_mapping_schematic}
\end{figure}
\begin{figure}[t]
    \centering
    \begin{minipage}[t]{0.49\columnwidth}
        \centering
        \includegraphics[width=0.6\columnwidth]{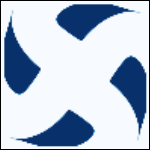}
        \subcaption{Optimized shape}
        \label{fig:example2_optimized_shape}
    \end{minipage}
    \begin{minipage}[t]{0.49\columnwidth}
        \centering
        \includegraphics[width=0.6\columnwidth]{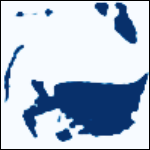}
        \subcaption{No symmetric shape}
        \label{fig:example2_no_symmetric_shape}
    \end{minipage}
    \caption{Optimized shape and optimized shape obtained without the cyclic periodic condition for the \textit{2D turbine}.}
\end{figure}
\begin{figure}[t]
    \centering
    \begin{minipage}[t]{0.65\columnwidth}
        \centering
        \includegraphics[width=\columnwidth]{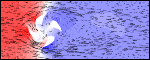}
        \subcaption{$t=41000\Delta t$}
    \end{minipage}
    \begin{minipage}[t]{0.07\columnwidth}
        \centering
        \includegraphics[width=\columnwidth]{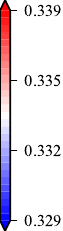}
    \end{minipage} \\
    \begin{minipage}[t]{0.65\columnwidth}
        \centering
        \includegraphics[width=\columnwidth]{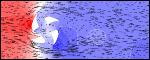}
        \subcaption{$t=43000\Delta t$}
    \end{minipage}
    \begin{minipage}[t]{0.07\columnwidth}
        \centering
        \includegraphics[width=\columnwidth]{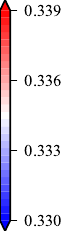}
    \end{minipage} \\
    \begin{minipage}[t]{0.65\columnwidth}
        \centering
        \includegraphics[width=\columnwidth]{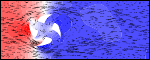}
        \subcaption{$t=45000\Delta t$}
    \end{minipage}
    \begin{minipage}[t]{0.07\columnwidth}
        \centering
        \includegraphics[width=\columnwidth]{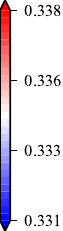}
    \end{minipage} \\
    \caption{Pressure distributions and flow velocity vector plots at each time step for the \textit{2D turbine}}
    \label{fig:example2_pressure}
\end{figure}
\begin{figure}[t]
    \centering
    \begin{minipage}[t]{0.49\columnwidth}
        \centering
        \includegraphics[width=0.6\columnwidth]{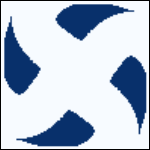}
        \subcaption{Binarized shape}
        \label{fig:example2_binarized_shape}
    \end{minipage}
    \begin{minipage}[t]{0.49\columnwidth}
        \centering
        \includegraphics[width=0.8\columnwidth]{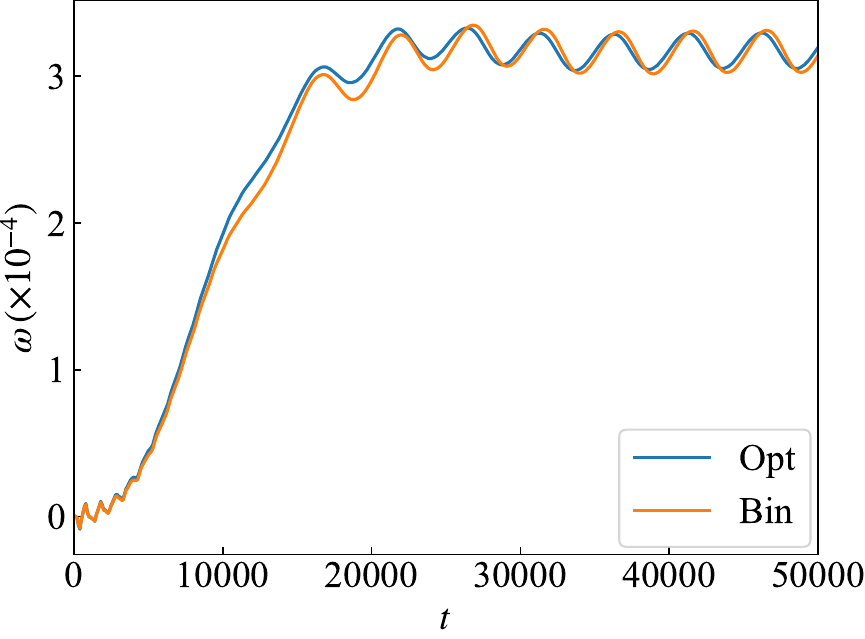}
        \subcaption{Comparison result}
        \label{fig:example2_check}
    \end{minipage}
    \caption{The binarized optimized shape and comparison result of the motion for the \textit{2D turbine}}
\end{figure}

As the second numerical example, we present the optimization of a rotating rigid-body shape driven by fluid flow force, referred to as the \textit{2D turbine}.
The design setting is shown in Fig.~\ref{fig:example2_schematic}.
In this example, the rigid body is allowed to undergo only rotational motion, and any translational motion is prohibited; in other words, only Eqs.~\eqref{eq:equation_of_motion_omega} and \eqref{eq:equation_of_motion_rotation} are considered.
The objective functional is the rotation angle given in Eq.~\eqref{eq:objective2}, whereas the constraint is minimum solid volume limitation given in Eq.~\eqref{eq:constraint}.
Unlike the previous example, a cyclic periodic shape, the procedure of which is illustrated in Fig.~\ref{fig:example2_mapping_schematic}, is applied in this example.
As shown in Fig.~\ref{fig:example2_mapping_schematic}, the design variables are defined on the left grid with four points colored magenta, green, yellow and blue.
Before computing the state and adjoint fields, the design variables are expanded in a cyclic periodic manner to the right grid with eight points.
In contrast, the design sensitivities computed on the right grid are collected back onto the left grid.
The parameters are as follows: $N_\text{t}=6000\Delta t$, $N_\text{opt}=40$, $N_\text{itr}=5$, $\nu=0.1$, $I_z=10^7$, $\Delta p=5\times10^{-3}$, $\kappa^\text{ref}_\text{max}=500$, $q=\left\{0.01,0.01,0.1,0.1,1.0,1.0\right\}$, $\beta=\left\{1,2,4,8,16,32\right\}$ and $V_\text{min}=0.25$.
In this example, the periodic problem approximation described in Section~\ref{sec34} is employed.

The optimized shape is shown in Fig.~\ref{fig:example2_optimized_shape}, whereas the optimized shape obtained without the cyclic periodic condition shown in Fig.~\ref{fig:example2_mapping_schematic} is presented in Fig.~\ref{fig:example2_no_symmetric_shape}.
In addition, the pressure distributions together with the flow velocity vector plots at each time step are shown in Fig.~\ref{fig:example2_pressure}.
It should be noted that, as stated above, the periodic problem approximation is employed in this example.
Accordingly, the optimization is performed only over a part of the time interval, with $N_\text{t}=6000\Delta t$, whereas the results shown in Fig.~\ref{fig:example2_pressure} are obtained from computations starting from the initial conditions, $\theta\left(0\right)=0$ and $\omega\left(0\right)=0$, using the optimized shape shown in Fig.~\ref{fig:example2_optimized_shape}.
The optimized shape consists of four parts resembling the leading edge of an airfoil, and the drag and lift forces acting on these parts generate rotational motion.
On the other hand, the optimized shape obtained without the cyclic periodic condition does not exhibit a meaningful structure.
This may be because, in this example, a periodic approximation approach is employed, and the shape is optimized for a given position at each optimization step; however, the position changes significantly in the subsequent step, which leads to instability in the optimization process.
The computational time per optimization step was approximately $350$~s.

As shown in Fig.~\ref{fig:example2_optimized_shape}, some grayscale regions remain, particularly near the trailing edge.
To evaluate the effect of such grayscale regions, we compare the motion of the optimized shape with that of the binarized shape.
The binarized shape with a threshold value of $0.5$ is shown in Fig.~\ref{fig:example2_binarized_shape}, and the comparison results are shown in Fig.~\ref{fig:example2_check}.
Here, ``Opt'' and ``Bin'' denote the results for the optimized shape and the binarized shape, respectively.
As in Fig.~\ref{fig:example2_pressure}, both results are computed from the initial conditions using each shape.
In Fig.~\ref{fig:example2_check}, the rotational ``velocity'' $\omega$, rather than the rotational ``position'' $\theta$, is shown, whereas in the previous example, Fig.~\ref{fig:example1_check}, the translational ``position'' $x^\text{G}_\alpha$ is shown.
This choice is made to clarify the comparison of the cyclic motion between the two shapes.
Although some discrepancies are observed, particularly in the transient phase (e.g., $t\leq25000\Delta t$), the overall trends are similar for both shapes.
Therefore, in practice, the approach of first optimizing a shape with some grayscale regions and subsequently binarizing it is effective.

\subsection{3D turbine}
\label{sec43}

\begin{figure}[t]
    \centering
    \begin{minipage}[t]{0.42\columnwidth}
        \centering
        \includegraphics[width=0.7\columnwidth]{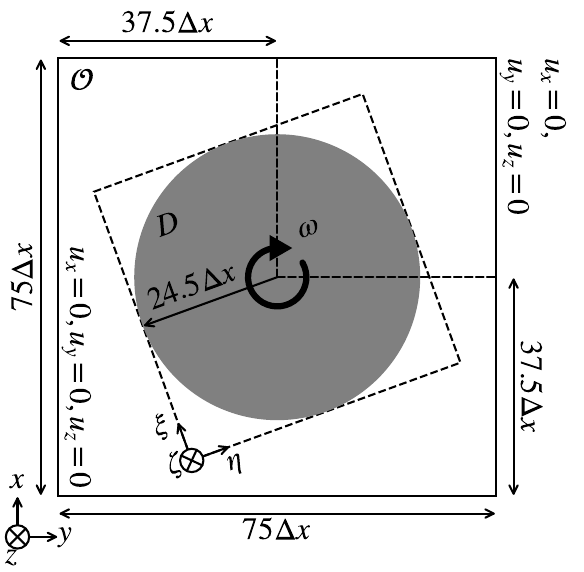}
        \subcaption{Front view}
    \end{minipage}
    \begin{minipage}[t]{0.56\columnwidth}
        \centering
        \includegraphics[width=0.7\columnwidth]{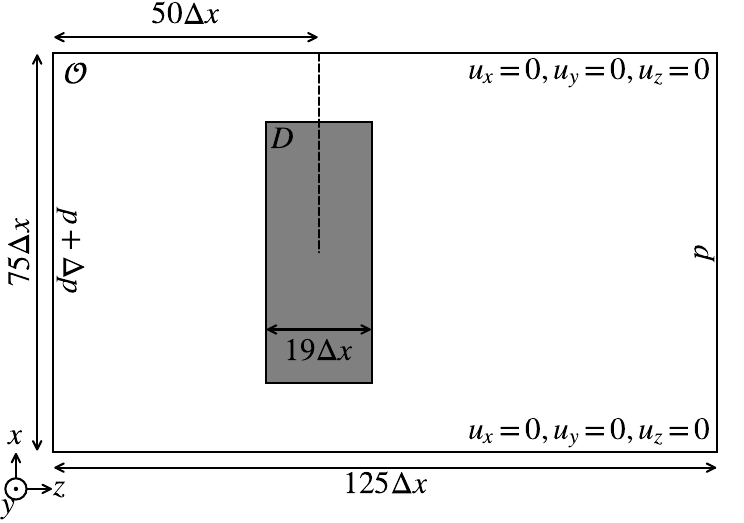}
        \subcaption{Side view}
    \end{minipage}
    \caption{Design setting of the \textit{3D turbine}.}
    \label{fig:example3_schematic}
\end{figure}
\begin{figure}[t]
    \centering
    \begin{minipage}[t]{0.29\columnwidth}
        \centering
        \includegraphics[width=\columnwidth]{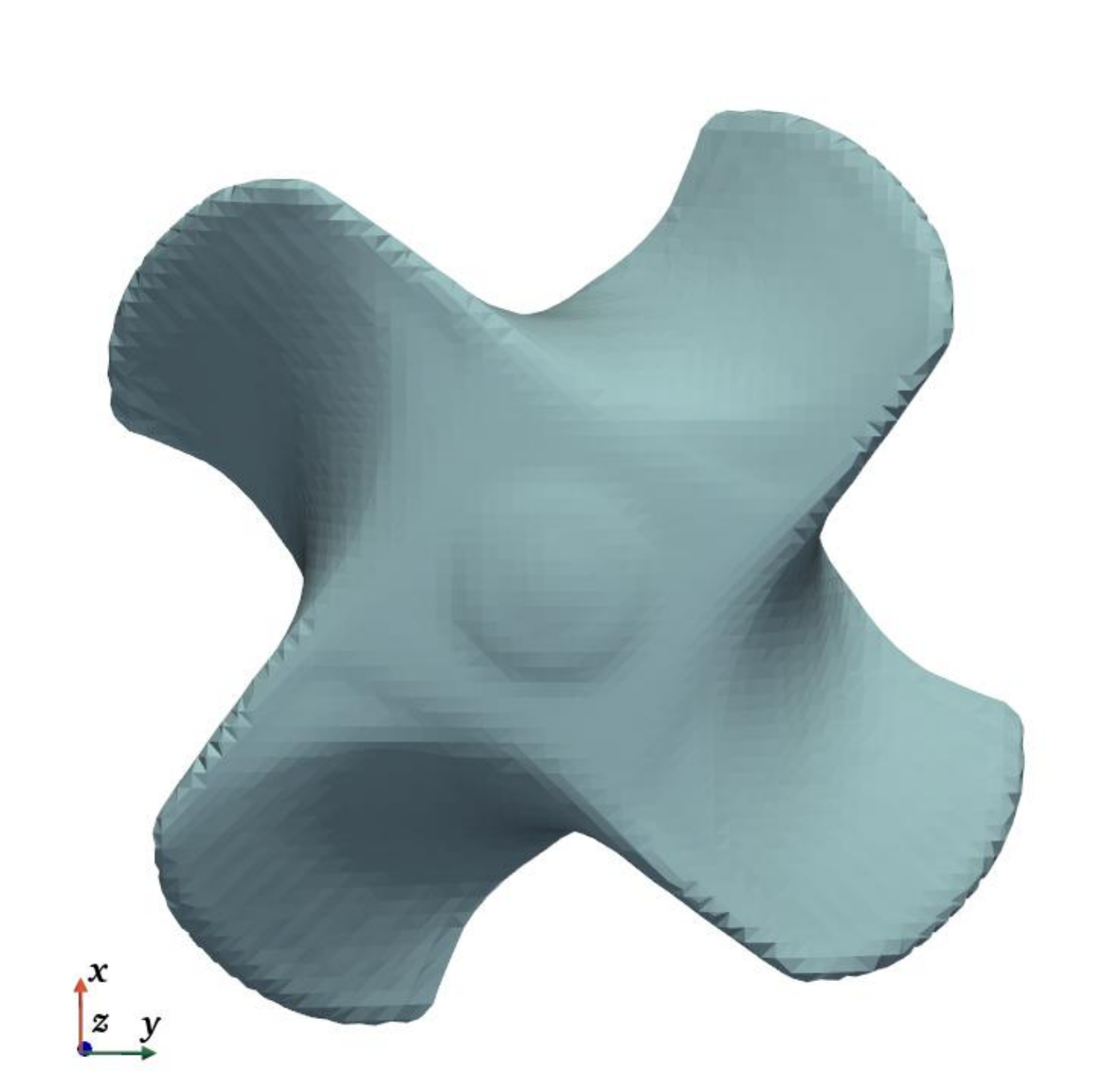}
        \subcaption{Front view}
    \end{minipage}
    \begin{minipage}[t]{0.29\columnwidth}
        \centering
        \includegraphics[width=0.65\columnwidth]{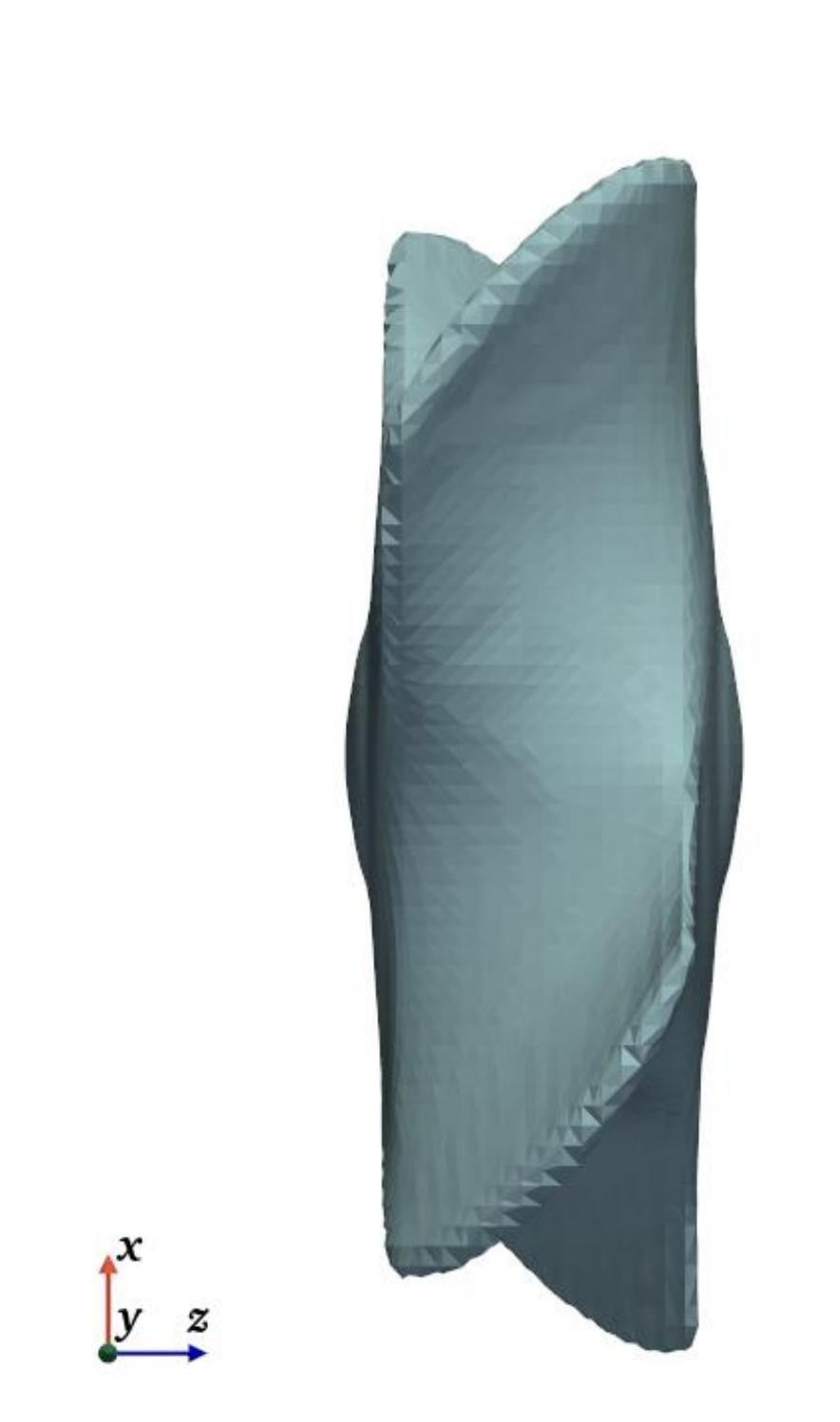}
        \subcaption{Side view}
    \end{minipage}
    \begin{minipage}[t]{0.38\columnwidth}
        \centering
        \includegraphics[width=\columnwidth]{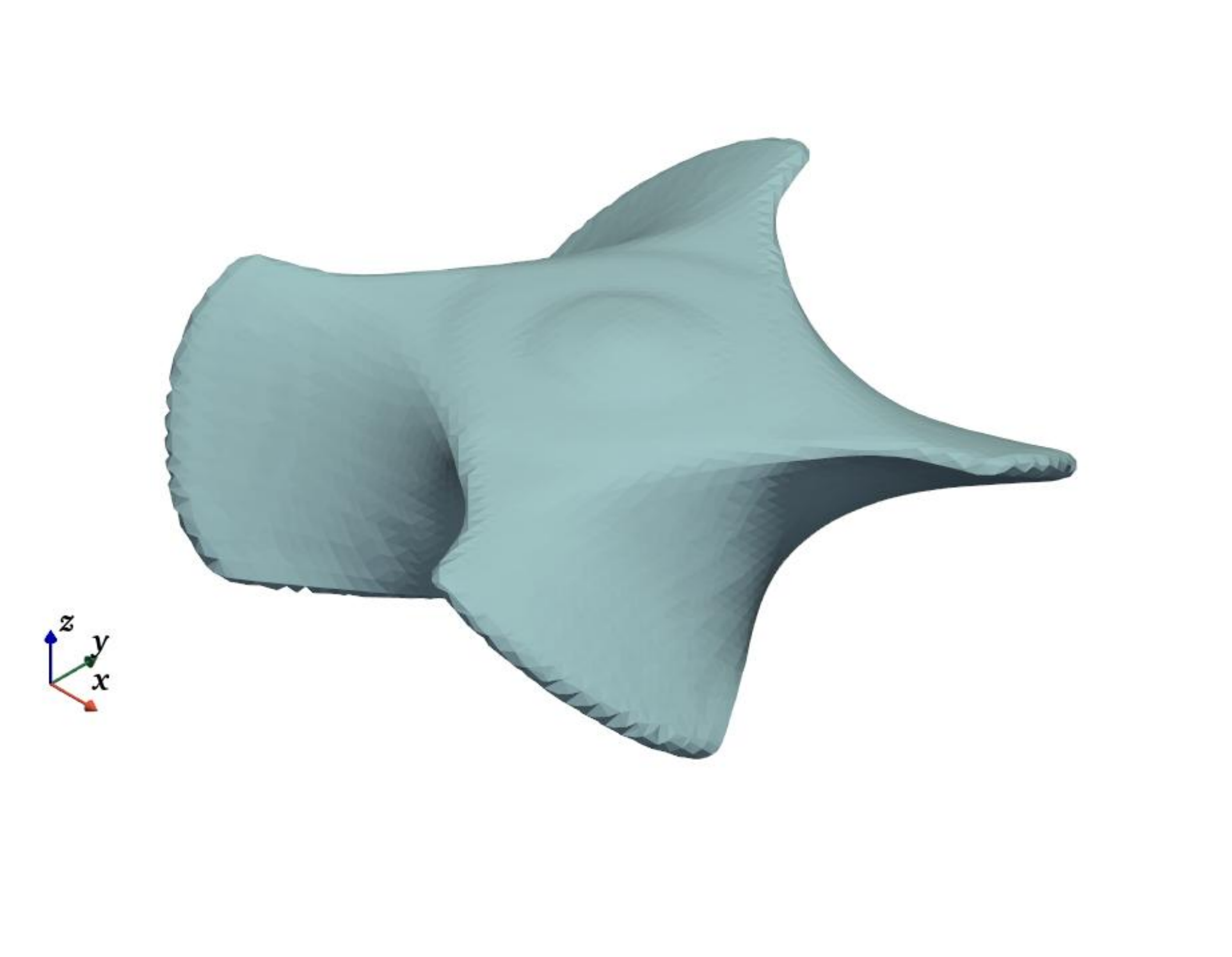}
        \subcaption{Isometric view}
    \end{minipage}
    \caption{Optimized shape of the \textit{3D turbine}}
    \label{fig:example3_optimized_shape}
\end{figure}
\begin{figure}[t]
    \centering
    \begin{minipage}[t]{0.4\columnwidth}
        \centering
        \includegraphics[width=\columnwidth]{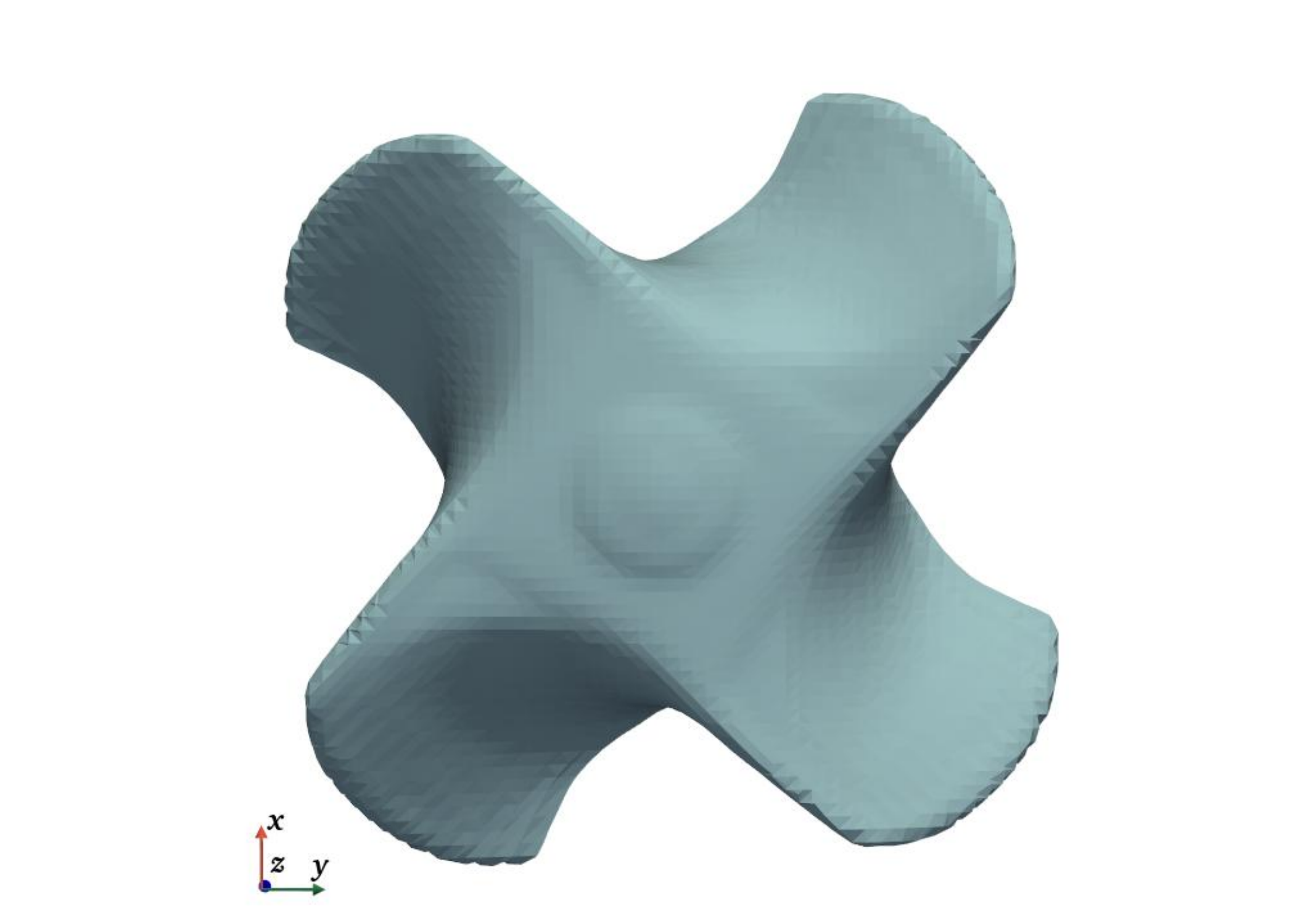}
    \end{minipage}
    \begin{minipage}[t]{0.44\columnwidth}
        \centering
        \includegraphics[width=\columnwidth]{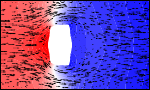}
        \subcaption{$t=7500\Delta t$}
    \end{minipage}
    \begin{minipage}[t]{0.1\columnwidth}
        \centering
        \includegraphics[width=\columnwidth]{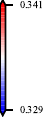}
    \end{minipage} \\
    \begin{minipage}[t]{0.4\columnwidth}
        \centering
        \includegraphics[width=\columnwidth]{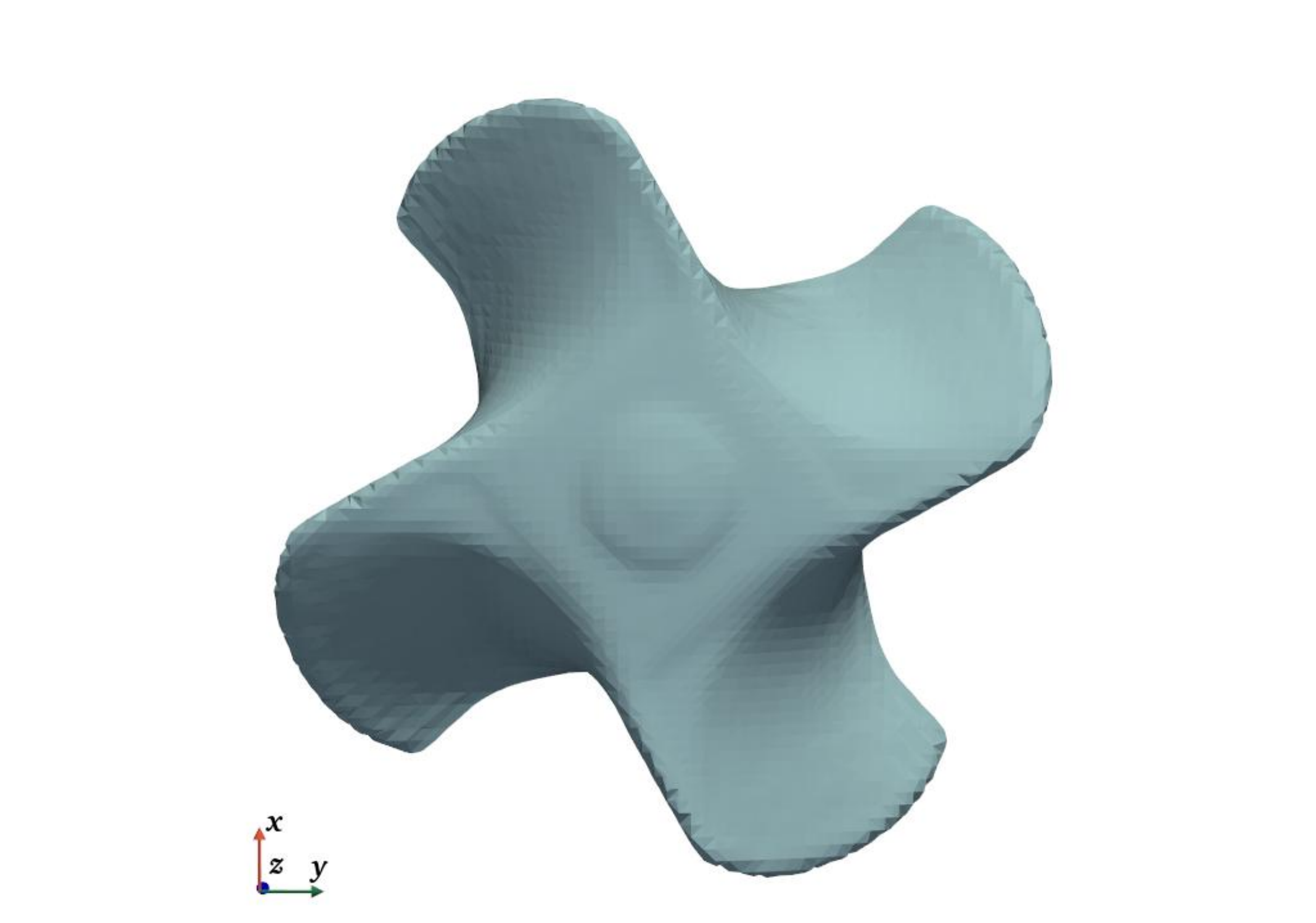}
    \end{minipage}
    \begin{minipage}[t]{0.44\columnwidth}
        \centering
        \includegraphics[width=\columnwidth]{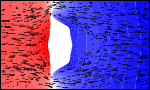}
        \subcaption{$t=7900\Delta t$}
    \end{minipage}
    \begin{minipage}[t]{0.1\columnwidth}
        \centering
        \includegraphics[width=\columnwidth]{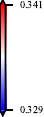}
    \end{minipage} \\
    \begin{minipage}[t]{0.4\columnwidth}
        \centering
        \includegraphics[width=\columnwidth]{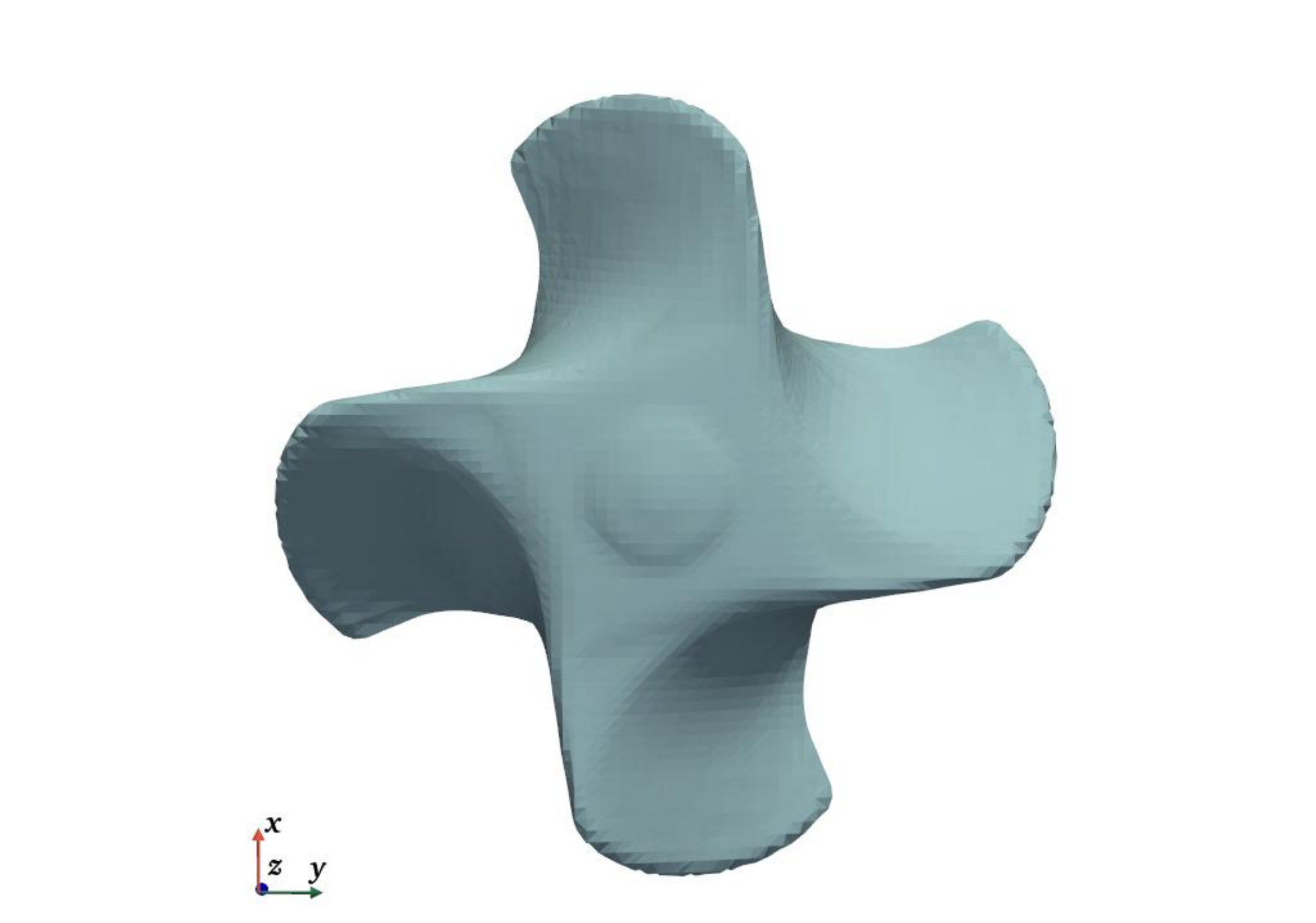}
    \end{minipage}
    \begin{minipage}[t]{0.44\columnwidth}
        \centering
        \includegraphics[width=\columnwidth]{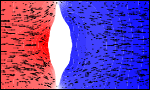}
        \subcaption{$t=8300\Delta t$}
    \end{minipage}
    \begin{minipage}[t]{0.1\columnwidth}
        \centering
        \includegraphics[width=\columnwidth]{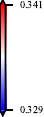}
    \end{minipage}
    \caption{Pressure distributions and flow velocity vector plots at each time step for the \textit{3D turbine}}
    \label{fig:example3_pressure}
\end{figure}
\begin{figure}[t]
    \centering
    \includegraphics[width=0.4\columnwidth]{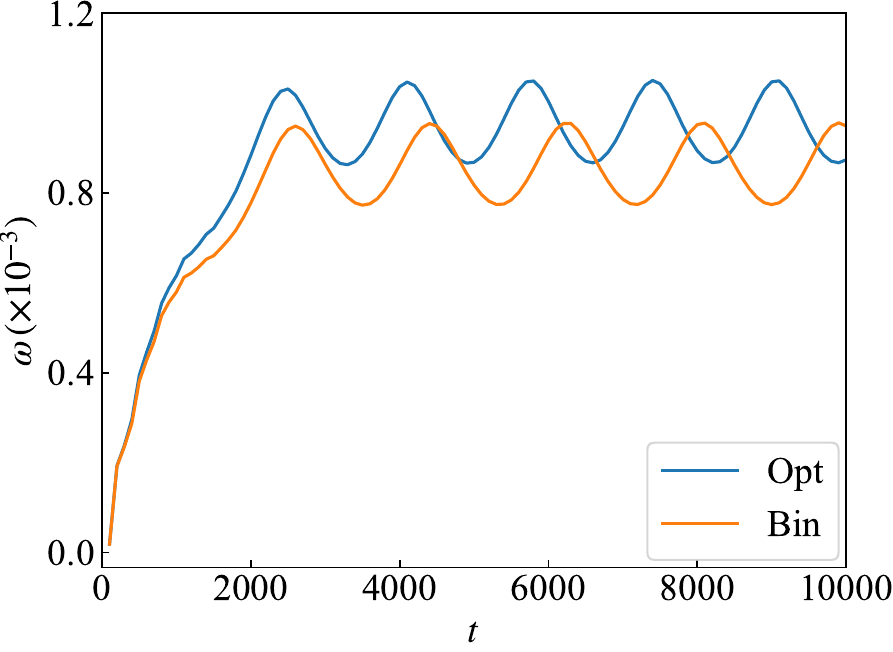}
    \caption{Comparison result of the motion for the \textit{3D turbine}.}
    \label{fig:example3_check}
\end{figure}

As the last example, we present the optimization of a rotating rigid-body shape in the three-dimensional case, referred to as the \textit{3D turbine}.
The design setting is shown in Fig.~\ref{fig:example3_schematic}.
As in the previous example, \textit{2D turbine}, the rigid body is allowed to undergo only rotational motion, and any translational motion is prohibited.
The objective functional is the rotation angle given in Eq.~\eqref{eq:objective2}, whereas the constraint is minimum solid volume limitation given in Eq.~\eqref{eq:constraint}.
In addition, a cyclic periodic condition is employed around the rotation axis.
The parameters are as follows: $N_\text{t}=2000\Delta t$, $N_\text{opt}=40$, $N_\text{itr}=5$, $\nu=0.1$, $I_z=10^7$, $\Delta p=10^{-2}$, $\kappa^\text{ref}_\text{max}=100$, $q=\left\{0.1,0.1,0.1,0.1,0.1\right\}$, $\beta=\left\{1,2,4,8,16\right\}$ and $V_\text{min}=0.25$.
The periodic approximation is also employed, as in the \textit{2D turbine}.

The optimized shape is shown in Fig.~\ref{fig:example3_optimized_shape}, and the pressure distributions together with the flow velocity vector plots at each time step are shown in Fig.~\ref{fig:example3_pressure}.
Here, the left column in Fig.~\ref{fig:example3_pressure} represents the front view of the rigid body at each time step.
As in the previous example, \textit{2D turbine}, the results in Fig.~\ref{fig:example3_pressure} is computed from the initial condition, $\theta\left(0\right)=0$ and $\omega\left(0\right)=0$, using the optimized shape.
The optimized shape resembles a turbine shape, and each blade receives fluid flow force, resulting in rotational motion.
The computational time per optimization step was approximately $1200$~s.

As in the previous two examples, some grayscale regions remain in the optimized shape.
Fig.~\ref{fig:example3_check} shows the comparison results between the optimized shape and the binarized shape.
Here, ``Opt'' and ``Bin'' denote the results for the optimized shape and the binarized shape, respectively.
Although the general trends---namely, a monotonic increase during the transient phase and periodical oscillations in the quasi-steady phase of the rotational velocity---are common to both shapes, non-negligible differences are observed in the phase, the period, and the time-averaged values.
In three-dimensional case, the mesh resolution is relatively coarse due to limitations in computational resources.
As a result, it is difficult to represent thin structures, such as turbine blades, without any grayscale regions, and discrepancies arise between the results with and without binarization. 

\section{Conclusion}
\label{sec5}

In this study, we propose a topology optimization method for moving rigid bodies driven by fluid flow forces.
The rigid-body motion is governed by the equations of motion, which are solved together with the governing equations of fluid flow in a two-way coupled manner.
The rigid body is represented using separated design-analysis grids, and the fluid flow is computed by the Lattice Kinetic Scheme (LKS), an extended version of the Lattice Boltzmann Method (LBM).
Design sensitivities are derived using the continuous adjoint variable method.
By incorporating several technics, such as iterative solution of fluid-rigid-body interaction, a filtering scheme, a continuation scheme, and a periodic approximation approach, the proposed method is applied to three optimization problems.
In the \textit{2D sail} example, a meaningful shape is obtained that resembles an airfoil and moves under lift forces, exhibiting better performance than the reference shape.
In the \textit{2D turbine} example, a suitable rotating shapes is obtained by employing a cyclic periodic configuration, and the effect of binarization---namely, the effect of grayscale regions---is negligible in practice.
In the \textit{3D turbine} example, a shape resembling a turbine is obtained; however, the difference due to binarization is not negligible, possibly because of the limited mesh resolution.
Improving the mesh resolution while saving computational cost in three-dimensional cases, such as by using an adaptive mesh method, and considering turbulent flows are directions for future work. 
\section{Replication of results}
The necessary information for replication of the results is presented in this paper.
The interested reader may contact the corresponding author for further implementation details.

\section*{Acknowledgements}
This work was supported by JSPS KAKENHI (GrantNo. 23K26018).

\appendix

\section{Derivation of the design sensitivity}
\label{appendixA}

We explain the derivation of the design sensitivity presented in Section~\ref{sec25}.
The functional derivatives of each term included in the Lagrangian $L$, defined in Eq.~\eqref{eq:lagrangian}, with respect to the design variable $\gamma$ are given as follows:
\begin{align}
    \langle J^\prime,\delta\gamma\rangle= & \int_\mathcal{I}\int_\mathcal{O}\frac{\partial J_\mathcal{O}}{\partial\gamma}\delta\gamma d\Omega dt+\int_\mathcal{I}\int_{\partial\mathcal{O}}\frac{\partial J_{\partial\mathcal{O}}}{\partial\gamma}\delta\gamma d\Gamma dt \notag                                                                                                                                  \\
                                          & +\int_\mathcal{I}\int_\mathcal{O}\sum_{i=0}^{Q-1}\frac{\partial J_\mathcal{O}}{\partial f_i}\delta f_id\Omega dt+\int_\mathcal{I}\int_{\partial\mathcal{O}}\sum_{i=0}^{Q-1}\frac{\partial J_{\partial\mathcal{O}}}{\partial f_i}\delta f_id\Gamma dt \notag                                                                                                           \\
                                          & +\int_\mathcal{I}\frac{\partial J_\mathcal{I}}{\partial x^\text{G}_\alpha}\delta x^\text{G}_\alpha dt+\int_\mathcal{I}\frac{\partial J_\mathcal{I}}{\partial u^\text{G}_\alpha}\delta u^\text{G}_\alpha dt+\int_\mathcal{I}\frac{\partial J_\mathcal{I}}{\partial\theta}\delta\theta dt+\int_\mathcal{I}\frac{\partial J_\mathcal{I}}{\partial\omega}\delta\omega dt,
\end{align}
\begin{align}
    \langle R_1^\prime,\delta\gamma\rangle= & \int_\mathcal{O}\text{Sh}\left[\tilde{f}_i\delta f_i\right]_{t_0}^{t_1}d\Omega+\int_\mathcal{I}\int_{\partial\mathcal{O}}\sum_in_\alpha\left\{\tilde{f}_i\delta_{\alpha\beta}-A\left(\tilde{s}_{\alpha\beta}+\tilde{s}_{\beta\alpha}\right)\right\}\delta f_id\Gamma dt \notag                                                                                                \\
                                            & +\int_\mathcal{I}\int_\mathcal{O}\sum_i\left\{-\text{Sh}\frac{\partial\tilde{f}_i}{\partial t}-c_{i\alpha}\frac{\partial\tilde{f}_i}{\partial x_\alpha}+\frac{1}{\varepsilon}\left(\tilde{f}_i-\tilde{f}_i^\text{eq}\right)+3\Delta xc_{i\alpha}\kappa\tilde{u}_\alpha\right\}\delta f_id\Omega dt \notag                                                                     \\
                                            & +\int_\mathcal{I}\int_\mathcal{O}\int_D3\Delta x\frac{\partial\kappa^\text{ref}}{\partial\gamma}W\left(u_\alpha-u^\text{S}_\alpha\right)\tilde{u}_\alpha\delta\gamma d\Omega^\text{ref}d\Omega dt \notag                                                                                                                                                                      \\
                                            & +\int_\mathcal{I}\int_\mathcal{O}\int_D3\Delta x\kappa^\text{ref}\frac{\partial W}{\partial x^\text{ref}_\alpha}\left(\frac{\partial x^\text{ref}_\alpha}{\partial x^\text{G}_\beta}\delta x^\text{G}_\beta+\frac{\partial x^\text{ref}_\alpha}{\partial\theta}\delta\theta\right)\left(u_\gamma-u^\text{S}_\gamma\right)\tilde{u}_\gamma d\Omega^\text{ref}d\Omega dt \notag \\
                                            & -\int_\mathcal{I}\int_\mathcal{O}\int_D3\Delta x\kappa\tilde{u}_\alpha\left(\frac{\partial u^\text{S,ref}_\alpha}{\partial u^\text{G}_\beta}\delta u^\text{G}_\beta+\frac{\partial u^\text{S,ref}_\alpha}{\partial\theta}\delta\theta+\frac{\partial u^\text{S,ref}_\alpha}{\partial\omega}\delta\omega\right)Wd\Omega^\text{ref}d\Omega dt \notag                            \\
                                            & -\int_\mathcal{I}\int_\mathcal{O}\int_D3\Delta x\kappa\tilde{u}_\alpha u^\text{S,ref}_\alpha\frac{\partial W}{\partial x^\text{ref}_\beta}\left(\frac{\partial x^\text{ref}_\beta}{\partial x^\text{G}_\gamma}\delta x^\text{G}_\gamma+\frac{\partial x^\text{ref}_\beta}{\partial\theta}\delta\theta\right)d\Omega^\text{ref}d\Omega dt, \label{eq:derivative_r1}
\end{align}
\begin{equation}
    \langle R_2^\prime,\delta\gamma\rangle=\left[\tilde{x}^\text{G}_\alpha\delta x^\text{G}_\alpha\right]_{t_0}^{t_1}-\int_\mathcal{I}\frac{d\tilde{x}^\text{G}_\alpha}{dt}\delta x^\text{G}_\alpha dt-\int_\mathcal{I}\tilde{x}^\text{G}_\alpha\delta u^\text{G}_\alpha dt,
\end{equation}
\begin{align}
    \langle R_3^\prime,\delta\gamma\rangle= & M\left[\tilde{u}^\text{G}_\alpha\delta u^\text{G}_\alpha\right]_{t_0}^{t_1}-\int_\mathcal{I}M\frac{d\tilde{u}^\text{G}_\alpha}{dt}\delta u^\text{G}_\alpha dt-\int_\mathcal{I}\tilde{u}^\text{G}_\alpha\int_D\frac{\partial\kappa^\text{ref}}{\partial\gamma}\delta\gamma\left(u^\text{ref}_\alpha-u^\text{S,ref}_\alpha\right)d\Omega^\text{ref}dt \notag \\
                                            & -\int_\mathcal{I}\tilde{u}^\text{G}_\alpha\int_D\int_\mathcal{O}\kappa^\text{ref}\delta u_\alpha Wd\Omega d\Omega^\text{ref}dt \notag                                                                                                                                                                                                                      \\
                                            & -\int_\mathcal{I}\tilde{u}^\text{G}_\alpha\int_D\int_\mathcal{O}\kappa^\text{ref}u_\alpha\frac{\partial W}{\partial x^\text{ref}_\beta}\left(\frac{\partial x^\text{ref}_\beta}{\partial x^\text{G}_\gamma}\delta x^\text{G}_\gamma+\frac{\partial x^\text{ref}_\beta}{\partial\theta}\delta\theta\right)d\Omega d\Omega^\text{ref}dt \notag               \\
                                            & +\int_\mathcal{I}\tilde{u}^\text{G}_\alpha\int_D\kappa^\text{ref}\left(\frac{\partial u^\text{S,ref}_\alpha}{\partial u^\text{G}_\beta}\delta u^\text{G}_\beta+\frac{\partial u^\text{S,ref}_\alpha}{\partial\theta}\delta\theta+\frac{\partial u^\text{S,ref}}{\partial\omega}\delta\omega\right)d\Omega^\text{ref}dt,
\end{align}
\begin{equation}
    \langle R_4^\prime,\delta\gamma\rangle=\left[\tilde{\theta}\delta\theta\right]_{t_0}^{t_1}-\int_\mathcal{I}\frac{d\tilde{\theta}}{dt}\delta\theta dt-\int_\mathcal{I}\tilde{\theta}\delta\omega dt,
\end{equation}
\begin{align}
    \langle R_5^\prime,\delta\gamma\rangle= & I_z\left[\tilde{\omega}\delta\omega\right]_{t_0}^{t_1}-\int_\mathcal{I}I_z\frac{d\tilde{\omega}}{dt}\delta\omega dt+\int_\mathcal{I}\tilde{\omega}\int_De_{z\alpha\beta}\delta x^\text{G}_\alpha\kappa^\text{ref}\left(u^\text{ref}_\beta-u^\text{S,ref}_\beta\right)d\Omega^\text{ref}dt \notag                                                                                                              \\
                                            & -\int_\mathcal{I}\tilde{\omega}\int_De_{z\alpha\beta}\left(\frac{\partial x^\text{ref}_\alpha}{\partial x^\text{G}_\gamma}\delta x^\text{G}_\gamma+\frac{\partial x^\text{ref}_\alpha}{\partial\theta}\delta\theta\right)\kappa^\text{ref}\left(u^\text{ref}_\beta-u^\text{S,ref}_\beta\right)d\Omega^\text{ref}dt \notag                                                                                     \\
                                            & -\int_\mathcal{I}\tilde{\omega}\int_De_{z\alpha\beta}\left(x^\text{ref}_\alpha-x^\text{G}_\alpha\right)\frac{\partial\kappa^\text{ref}}{\partial\gamma}\delta\gamma\left(u^\text{ref}_\beta-u^\text{S,ref}_\beta\right)d\Omega^\text{ref}dt \notag                                                                                                                                                            \\
                                            & -\int_\mathcal{I}\tilde{\omega}\int_D\int_\mathcal{O}e_{z\alpha\beta}\left(x^\text{ref}_\alpha-x^\text{G}_\alpha\right)\kappa^\text{ref}\delta u_\beta Wd\Omega d\Omega^\text{ref}dt \notag                                                                                                                                                                                                                   \\
                                            & -\int_\mathcal{I}\tilde{\omega}\int_D\int_\mathcal{O}e_{z\alpha\beta}\left(x^\text{ref}_\alpha-x^\text{G}_\alpha\right)\kappa^\text{ref}u_\beta\frac{\partial W}{\partial x^\text{ref}_\gamma}\left(\frac{\partial x^\text{ref}_\gamma}{\partial x^\text{G}_\delta}\delta x^\text{G}_\delta+\frac{\partial x^\text{ref}_\gamma}{\partial\theta}\delta\theta\right)d\Omega d\Omega^\text{ref}dt \notag         \\
                                            & +\int_\mathcal{I}\tilde{\omega}\int_De_{z\alpha\beta}\left(x^\text{ref}_\alpha-x^\text{G}_\alpha\right)\kappa^\text{ref}\left(\frac{\partial u^\text{S,ref}_\beta}{\partial u^\text{G}_\gamma}\delta u^\text{G}_\gamma+\frac{\partial u^\text{S,ref}_\beta}{\partial\theta}\delta\theta+\frac{\partial u^\text{S,ref}_\beta}{\partial\omega}\delta\omega\right)d\Omega^\text{ref}dt. \label{eq:derivative_r5}
\end{align}
According to Eq.~\eqref{eq:rigid_body_position}, the derivatives of $x^\text{ref}_\alpha$ with respect to $x^\text{G}_\beta$ and $\theta$ are given as follows:
\begin{align}
     & \frac{\partial x^\text{ref}_\alpha}{\partial x^\text{G}_\beta}=\delta_{\alpha\beta},                                                                               \\
     & \left(\begin{array}{c}
                     \partial x^\text{ref}_x/\partial\theta \\
                     \partial x^\text{ref}_y/\partial\theta
                 \end{array}\right)=  \left(\begin{array}{cc}
                                                -\sin{\theta\left(t\right)} & -\cos{\theta\left(t\right)} \\
                                                \cos{\theta\left(t\right)}  & -\sin{\theta\left(t\right)}
                                            \end{array}\right)\left(\begin{array}{c}\xi-\xi^{\text{G}}\\\eta-\eta^{\text{G}}\end{array}\right).
\end{align}
Similarly, based on Eq.~\eqref{eq:rigid_body_velocity}, the derivatives of $u^\text{S,ref}_\alpha$ with respect to $u^\text{G}_\beta$, $\theta$ and $\omega$ are given as follows:
\begin{align}
     & \frac{\partial u^\text{S,ref}_\alpha}{\partial u^\text{G}_\beta}=\delta_{\alpha\beta},                                                                                               \\
     & \left(\begin{array}{c}
                     \partial u^\text{S,ref}_x/\partial\theta \\
                     \partial u^\text{S,ref}_y/\partial\theta
                 \end{array}\right)=\left(\begin{array}{cc}
                                              -\cos{\theta\left(t\right)} & \sin{\theta\left(t\right)}  \\
                                              -\sin{\theta\left(t\right)} & -\cos{\theta\left(t\right)}
                                          \end{array}\right)\left(\begin{array}{c}\xi-\xi^{\text{G}}\\\eta-\eta^{\text{G}}\end{array}\right)\omega\left(t\right), \\
     & \left(\begin{array}{c}
                     \partial u^\text{S,ref}_x/\partial\omega \\
                     \partial u^\text{S,ref}_y/\partial\omega
                 \end{array}\right)=\left(\begin{array}{cc}
                                              -\sin{\theta\left(t\right)} & -\cos{\theta\left(t\right)} \\
                                              \cos{\theta\left(t\right)}  & -\sin{\theta\left(t\right)}
                                          \end{array}\right)\left(\begin{array}{c}\xi-\xi^{\text{G}}\\\eta-\eta^{\text{G}}\end{array}\right).
\end{align}

To avoid explicitly computing the derivatives of the state fields with respect to the design variable, Eqs.~\eqref{eq:derivative_r1}--\eqref{eq:derivative_r5} are rearranged such that the coefficients of $\delta f_i$, $\delta x^\text{G}_\alpha$, $\delta u^\text{G}_\alpha$, $\delta\theta$, and $\delta\omega$ vanish.
From the coefficients of $\delta f_i$, the adjoint equations are obtained as follows:
\begin{align}
     & \left.\tilde{f}_i\right|_{t=t_1}=                                                                                                                                           -\left.\frac{\partial J_\mathcal{O}}{\partial f_i}\right|_{t=t_1},                                                                                                                                                                                                   \\
     & n_\alpha\left\{\tilde{f}_i\delta_{\alpha\beta}-A\left(\tilde{s}_{\alpha\beta}+\tilde{s}_{\beta\alpha}\right)\right\}+\frac{\partial J_{\partial\mathcal{O}}}{\partial f_i}= 0,                                                                                                                                                                                                                                                                   \\
     & \begin{aligned}
           -\text{Sh}\frac{\partial\tilde{f}_i}{\partial t}-c_{i\alpha}\frac{\partial\tilde{f}_i}{\partial x_\alpha}= & -\frac{1}{\varepsilon}\left(\tilde{f}_i-\tilde{f}_i^\text{eq}\right)-3\Delta xc_{i\alpha}\kappa\tilde{u}_\alpha                                                                                                                                                                                         \\
                                                                                                                      & +c_{i\alpha}\tilde{u}^\text{G}_\alpha\int_D\kappa^\text{ref}Wd\Omega^\text{ref}+c_{i\beta}\tilde{\omega}\int_De_{z\alpha\beta}\left(x^\text{ref}_\alpha-x^\text{G}_\alpha\right)\kappa^\text{ref}Wd\Omega^\text{ref}-\frac{\partial J_\mathcal{O}}{\partial f_i}, \label{eq:adjoint_discrete_boltzmann}
       \end{aligned}
\end{align}
where $f^\text{eq}_i$ is defined as follows:
\begin{equation}
    \tilde{f}^\text{eq}_i=\tilde{p}+3c_{i\alpha}\left(\tilde{u}_\alpha+3\tilde{s}_{\alpha\beta}u_\beta-\tilde{p}u_\alpha\right)-\Delta xA\frac{\partial}{\partial x_\beta}\left(\tilde{s}_{\alpha\beta}+\tilde{s}_{\beta\alpha}\right)c_{i\alpha}.
\end{equation}
Here, $\tilde{p}$, $\tilde{u}_\alpha$, and $\tilde{s}_{\alpha\beta}$ are the macroscopic variables of the adjoint equations and are defined as the moments of $\tilde{f}_i$, as follows:
\begin{align}
     & \tilde{p}=\sum_{i=0}^{Q-1}w_i\tilde{f}_i, \label{eq:ap_continuous_form}                                   \\
     & \tilde{u}_\alpha=\sum_{i=0}^{Q-1}w_ic_{i\alpha}\tilde{f}_i, \label{eq:au_continuous_form}                 \\
     & \tilde{s}_{\alpha\beta}=\sum_{i=0}^{Q-1}w_ic_{i\alpha}c_{i\beta}\tilde{f}_i.\label{eq:as_continuous_form}
\end{align}
In the same manner as in the LKS, Eq.~\eqref{eq:adjoint_discrete_boltzmann} is discretized as follows:
\begin{align}
     & \tilde{f}^*_i\left(\bm{x}-\bm{c}_i\Delta x,t-\Delta t\right)=\tilde{f}_i\left(\bm{x},t\right)-\frac{1}{\tau}\left\{\tilde{f}_i\left(\bm{x},t\right)-\tilde{f}^\text{eq}_i\left(\bm{x},t\right)\right\},         \label{eq:alb_tmp}                                                                                                                                                                                                                                        \\
     & \begin{aligned}
           \tilde{f}_i\left(\bm{x},t\right)= & \tilde{f}^*_i\left(\bm{x},t\right)-3\Delta xc_{i\alpha}\kappa\left(\bm{x},t\right)\tilde{u}_\alpha\left(\bm{x},t\right)+c_{i\alpha}\tilde{u}^\text{G}_\alpha\left(t\right)\sum_{m=1}^{N_\text{grid}^\text{ref}}\kappa^\text{ref}\left(\bm{\xi}_m\right)W\left(\bm{x},\bm{x}^\text{ref}\left(\bm{\xi}_m,t\right)\right)                                                     \\
                                             & +c_{i\alpha}\tilde{\omega}\left(t\right)\sum_{m=1}^{N_\text{grid}^\text{ref}}e_{z\alpha\beta}\left\{x^\text{ref}_\alpha\left(\bm{\xi},t\right)-x^\text{G}_\alpha\left(t\right)\right\}\kappa^\text{ref}\left(\bm{\xi}\right)W\left(\bm{x},\bm{x}^\text{ref}\left(\bm{\xi},t\right)\right)-\frac{\partial J_\mathcal{O}}{\partial f_i}\left(\bm{x},t\right). \label{eq:alb}
       \end{aligned}
\end{align}
Here, in the LKS, $\tau=1$; therefore, Eq.~\eqref{eq:alb_tmp} is simplified as
\begin{equation}
    \tilde{f}^*_i\left(\bm{x}-\bm{c}_i\Delta x,t-\Delta t\right)=\tilde{f}^\text{eq}_i\left(\bm{x},t\right). \label{eq:af}
\end{equation}
Substituting Eq.~\eqref{eq:af} into Eqs.~\eqref{eq:ap_continuous_form}--\eqref{eq:as_continuous_form}, Eqs.~\eqref{eq:ap_tmp}--\eqref{eq:as_tmp} are obtained.
By taking the moments of Eq.~\eqref{eq:alb}, Eqs.~\eqref{eq:ap}--\eqref{eq:as} are obtained.

In the same manner as for the fluid adjoint equations, the adjoint equations for rigid-body motion are derived as follows.
For $\delta x^\text{G}_\alpha$:
\begin{align}
    \tilde{x}^\text{G}_\alpha\left(t_1\right)= & -\left.\frac{\partial J_\mathcal{I}}{\partial x^\text{G}_\alpha}\right|_{t=t_1}                                                                                                                                                                                                                                                                                   \\
    -\frac{d\tilde{x}^\text{G}_\alpha}{dt}=    & -\int_\mathcal{O}\int_D3\Delta x\kappa^\text{ref}\frac{\partial W}{\partial x^\text{ref}_\beta}\frac{\partial x^\text{ref}_\beta}{\partial x^\text{G}_\alpha}\left(u_\gamma-u^\text{S}_\gamma\right)\tilde{u}_\gamma d\Omega^\text{ref}d\Omega \notag                                                                                                             \\
                                               & +\int_\mathcal{O}\int_D3\Delta x\kappa\tilde{u}_\gamma u^\text{S,ref}_\gamma\frac{\partial W}{\partial x^\text{ref}_\beta}\frac{\partial x^\text{ref}_\beta}{\partial x^\text{G}_\alpha}d\Omega^\text{ref}d\Omega \notag                                                                                                                                          \\
                                               & +\tilde{u}^\text{G}_\gamma\int_D\int_\mathcal{O}\kappa^\text{ref}u_\gamma\frac{\partial W}{\partial x^\text{ref}_\beta}\frac{\partial x^\text{ref}_\beta}{\partial x^\text{G}_\alpha}d\Omega d\Omega^\text{ref} \notag                                                                                                                                            \\
                                               & -\tilde{\omega}\int_De_{z\alpha\beta}\kappa^\text{ref}\left(u^\text{ref}_\beta-u^\text{S,ref}_\beta\right)d\Omega^\text{ref}+\tilde{\omega}\int_De_{z\gamma\beta}\frac{\partial x^\text{ref}_\gamma}{\partial x^\text{G}_\alpha}\kappa^\text{ref}\left(u^\text{ref}_\beta-u^\text{S,ref}_\beta\right)d\Omega^\text{ref} \notag                                    \\
                                               & +\tilde{\omega}\int_D\int_\mathcal{O}e_{z\delta\beta}\left(x^\text{ref}_\delta-x^\text{G}_\delta\right)\kappa^\text{ref}u_\beta\frac{\partial W}{\partial x^\text{ref}_\gamma}\frac{\partial x^\text{ref}_\gamma}{\partial x^\text{G}_\alpha}d\Omega d\Omega^\text{ref}-\frac{\partial J_\mathcal{I}}{\partial x^\text{G}_\alpha}, \label{eq:axg_continuous_form}
\end{align}
For $\delta u^\text{G}_\alpha$:
\begin{align}
    M\tilde{u}^\text{G}_\alpha\left(t_1\right)= & -\left.\frac{\partial J_\mathcal{I}}{\partial u^\text{G}_\alpha}\right|_{t=t_1}                                                                                                                                                                                                                                   \\
    -M\frac{d\tilde{u}^\text{G}_\alpha}{dt}=    & \tilde{x}^\text{G}_\alpha+\int_\mathcal{O}\int_D3\Delta x\kappa\tilde{u}_\beta\frac{\partial u^\text{S,ref}_\beta}{\partial u^\text{G}_\alpha}Wd\Omega^\text{ref}d\Omega-\tilde{u}^\text{G}_\beta\int_D\kappa^\text{ref}\frac{\partial u^\text{S,ref}_\beta}{\partial u^\text{G}_\alpha}d\Omega^\text{ref} \notag \\
                                                & -\tilde{\omega}\int_De_{z\gamma\beta}\left(x^\text{ref}_\gamma-x^\text{G}_\gamma\right)\kappa^\text{ref}\frac{\partial u^\text{S,ref}_\beta}{\partial u^\text{G}_\alpha}d\Omega^\text{ref}-\frac{\partial J_\mathcal{I}}{\partial u^\text{G}_\alpha}, \label{eq:aug_continuous_form}
\end{align}
For $\delta\theta$:
\begin{align}
    \tilde{\theta}\left(t_1\right)= & -\left.\frac{\partial J_\mathcal{I}}{\partial\theta}\right|_{t=t_1}                                                                                                                                                                                                                                                                               \\
    -\frac{d\tilde{\theta}}{dt}=    & -\int_\mathcal{O}\int_D3\Delta x\kappa^\text{ref}\frac{\partial W}{\partial x^\text{ref}_\alpha}\frac{\partial x^\text{ref}_\alpha}{\partial\theta}\left(u_\gamma-u^\text{S}_\gamma\right)\tilde{u}_\gamma d\Omega^\text{ref}d\Omega \notag                                                                                                       \\
                                    & +\int_\mathcal{O}\int_D3\Delta x\kappa\tilde{u}_\alpha\frac{\partial u^\text{S,ref}_\alpha}{\partial\theta}Wd\Omega^\text{ref}d\Omega+\int_\mathcal{O}\int_D3\Delta x\kappa\tilde{u}_\alpha u^\text{S,ref}_\alpha\frac{\partial W}{\partial x^\text{ref}_\beta}\frac{\partial x^\text{ref}_\beta}{\partial\theta}d\Omega^\text{ref}d\Omega \notag \\
                                    & +\tilde{u}^\text{G}_\alpha\int_D\int_\mathcal{O}\kappa^\text{ref}u_\alpha\frac{\partial W}{\partial x^\text{ref}_\beta}\frac{\partial x^\text{ref}_\beta}{\partial\theta}d\Omega d\Omega^\text{ref}-\tilde{u}^\text{G}_\alpha\int_D\kappa^\text{ref}\frac{\partial u^\text{S,ref}_\alpha}{\partial\theta}d\Omega^\text{ref}\notag                 \\
                                    & +\tilde{\omega}\int_De_{z\alpha\beta}\frac{\partial x^\text{ref}_\alpha}{\partial\theta}\kappa^\text{ref}\left(u^\text{ref}_\beta-u^\text{S,ref}_\beta\right)d\Omega^\text{ref} \notag                                                                                                                                                            \\
                                    & +\tilde{\omega}\int_D\int_\mathcal{O}e_{z\alpha\beta}\left(x^\text{ref}_\alpha-x^\text{G}_\alpha\right)\kappa^\text{ref}u_\beta\frac{\partial W}{\partial x^\text{ref}_\gamma}\frac{\partial x^\text{ref}_\gamma}{\partial\theta}d\Omega d\Omega^\text{ref} \notag                                                                                \\
                                    & -\tilde{\omega}\int_De_{z\alpha\beta}\left(x^\text{ref}_\alpha-x^\text{G}_\alpha\right)\kappa^\text{ref}\frac{\partial u^\text{S,ref}_\beta}{\partial\theta}d\Omega^\text{ref}-\frac{\partial J_\mathcal{I}}{\partial\theta}, \label{eq:atheta_continuous_form}
\end{align}
For $\delta\omega$:
\begin{align}
    I_z\tilde{\omega}\left(t_1\right)= & -\left.\frac{\partial J_\mathcal{I}}{\partial\omega}\right|_{t=t_1},                                                                                                                                                                                                        \\
    -I_z\frac{d\tilde{\omega}}{dt}=    & \tilde{\theta}+\int_\mathcal{O}\int_D3\Delta x\kappa\tilde{u}_\alpha\frac{\partial u^\text{S,ref}_\alpha}{\partial\omega}Wd\Omega^\text{ref}d\Omega-\tilde{u}^\text{G}_\alpha\int_D\kappa^\text{ref}\frac{\partial u^\text{S,ref}}{\partial\omega}d\Omega^\text{ref} \notag \\
                                       & -\tilde{\omega}\int_De_{z\alpha\beta}\left(x^\text{ref}_\alpha-x^\text{G}_\alpha\right)\kappa^\text{ref}\frac{\partial u^\text{S,ref}_\beta}{\partial\omega}d\Omega^\text{ref}-\frac{\partial J_\mathcal{I}}{\partial\omega}, \label{eq:aomega_continuous_form}
\end{align}
The adjoint equations, Eqs.~\eqref{eq:axg}, \eqref{eq:aug}, \eqref{eq:atheta}, and \eqref{eq:aomega}, presented in Section~\ref{sec25}, are obtained by discretizing Eqs.~\eqref{eq:axg_continuous_form}, \eqref{eq:aug_continuous_form}, \eqref{eq:atheta_continuous_form}, and \eqref{eq:aomega_continuous_form} in the same manner as the equations of motion described in Section~\ref{sec24}.
Specifically, the time-derivative terms in each equation are discretized using the Euler method; for example,
\begin{equation}
    \frac{\partial\tilde{x}^\text{G}_\alpha}{\partial t}\simeq\frac{\tilde{x}^\text{G}_\alpha\left(t-\Delta t\right)-\tilde{x}^\text{G}_\alpha\left(t\right)}{\Delta t},
\end{equation}
and the terms on the right-hand side are evaluated at time $t-\Delta t$. 
\section{Verification of sensitivity analysis}
\label{appendixB}

In this section, we verify the sensitivity analysis through two test cases: one for translational motion and the other for rotational motion.
The results obtained by the proposed method are compared with those obtained by the finite difference approximation (FDA).

\begin{figure}[t]
    \centering
    \begin{minipage}[t]{0.49\columnwidth}
        \centering
        \includegraphics[width=0.7\columnwidth]{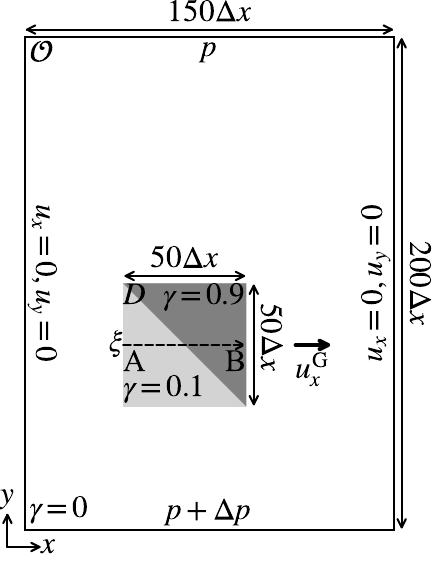}
        \subcaption{Problem setting}
        \label{fig:translation_sensitivity_setting}
    \end{minipage}
    \begin{minipage}[t]{0.49\columnwidth}
        \centering
        \includegraphics[width=\columnwidth]{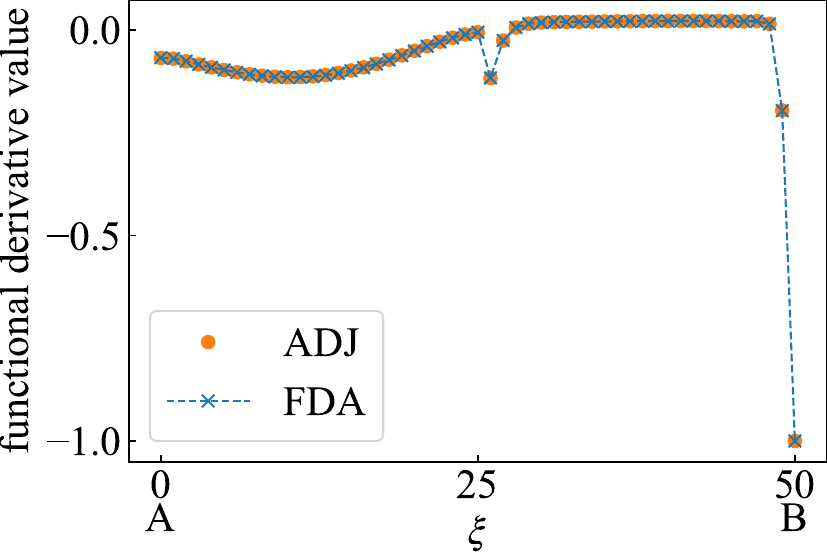}
        \subcaption{Comparison result}
        \label{fig:translational_sensitivity_result}
    \end{minipage}
    \caption{Problem setting and comparison results for the verification of the sensitivity analysis in the translational case.}
\end{figure}

First, the translational case is considered, and the problem setting is shown in Fig.~\ref{fig:translation_sensitivity_setting}.
Here, the rigid body is allowed to move only horizontally, while rotation and vertical translation are prohibited.
The compared functional is the translational distance of the rigid-body center of gravity defined as Eq.~\eqref{eq:objective1}, and the design sensitivity is compared along the dashed line in Fig.~\ref{fig:translation_sensitivity_setting}.
The parameters are as follows: $N_\text{t}=3000\Delta t$, $\nu=0.1$, $M=10^3$, $\Delta p=10^{-2}$, $\kappa^\text{ref}_\text{max}=300$, and $q=10^{-2}$.
The comparison results are shown in Fig.~\ref{fig:translational_sensitivity_result}, where ``ADJ'' denotes the result obtained by the proposed method.
As shown in Fig.~\ref{fig:translational_sensitivity_result}, the two results are in good agreement.

\begin{figure}[t]
    \centering
    \begin{minipage}[t]{0.49\columnwidth}
        \centering
        \includegraphics[width=0.7\columnwidth]{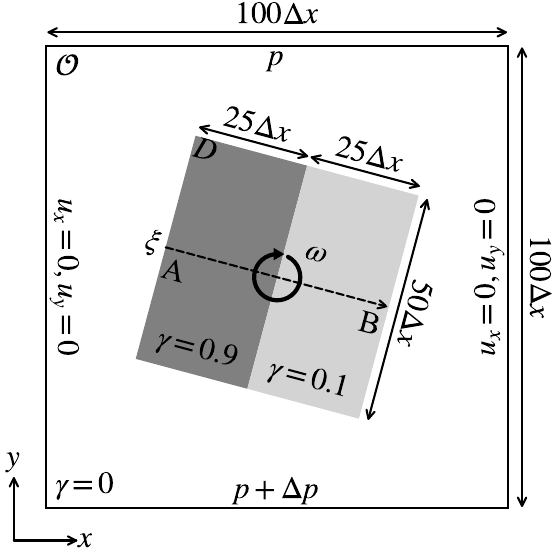}
        \subcaption{Problem setting}
        \label{fig:rotation_sensitivity_setting}
    \end{minipage}
    \begin{minipage}[t]{0.49\columnwidth}
        \centering
        \includegraphics[width=\columnwidth]{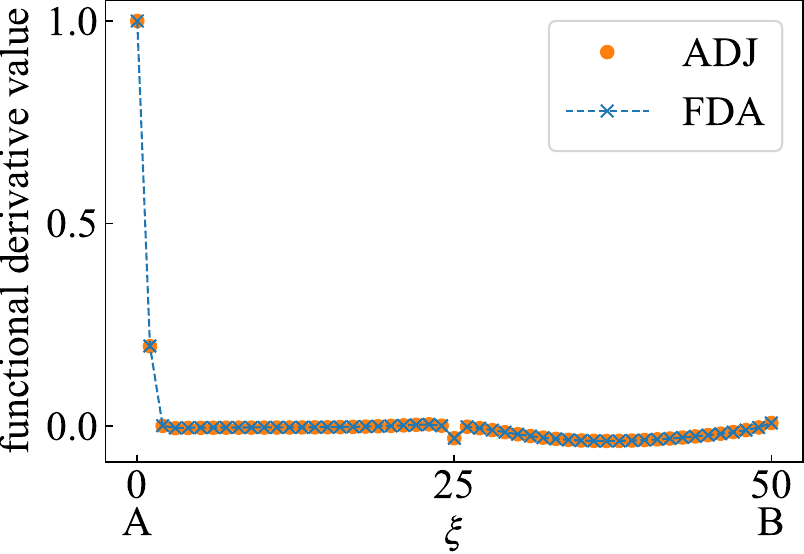}
        \subcaption{Comparison result}
        \label{fig:rotation_sensitivity_result}
    \end{minipage}
    \caption{Problem setting and comparison results for the verification of the sensitivity analysis in the rotational case.}
\end{figure}

Next, the rotational case is considered, and the problem setting is shown in Fig.~\ref{fig:rotation_sensitivity_setting}.
Here, the rigid body is allowed to undergo rotational motion, and any translational motion is prohibited.
The compared functional is the rotation angle of the rigid body defined as Eq.~\eqref{eq:objective2}, and the design sensitivity is compared along the dashed line in Fig.~\ref{fig:rotation_sensitivity_setting}.
The parameters are as follows: $N_\text{t}=10000\Delta t$, $\nu=0.1$, $I_z=10^6$, $\Delta p=10^{-2}$, $\kappa^\text{ref}_\text{max}=200$, and $q=10^{-2}$.
The comparison results are shown in Fig.~\ref{fig:rotation_sensitivity_result}, and the two results are in good agreement. 
\section*{Conflict of interest}
The authors declare that they have no conflict of interest.

\end{document}